\documentclass[twocolumn,floatfix,aps,prd,showpacs]{revtex4}

\usepackage{graphicx}
\usepackage{dcolumn}
\usepackage{bm}
\usepackage{epsf}

\draft

\newcommand{\beq}{\begin{equation}}
\newcommand{\eeq}{\end{equation}}
\newcommand{\beqn}{\begin{eqnarray}}
\newcommand{\eeqn}{\end{eqnarray}}

\def\agt{\mathrel{\raise.3ex\hbox{$>$}\mkern-14mu\lower0.6ex\hbox{$\sim$}}}
\def\alt{\mathrel{\raise.3ex\hbox{$<$}\mkern-14mu\lower0.6ex\hbox{$\sim$}}}

\begin{document}

\title{Stably stratified magnetized stars in general relativity} 

\author{Shijun Yoshida$^1$}
\author{Kenta Kiuchi$^2$}
\author{Masaru Shibata$^2$}

\affiliation{
$^{1}$Astronomical Institute, Tohoku University, Sendai 980-8578, Japan\\
$^{2}$Yukawa Institute for Theoretical Physics, Kyoto University, Kyoto 606-8502, Japan}

\begin{abstract}
We construct magnetized stars composed of a fluid stably stratified by
entropy gradients in the framework of general relativity, assuming
ideal magnetohydrodynamics and employing a barotropic equation of
state. We first revisit basic equations for describing
stably-stratified stationary axisymmetric stars containing both
poloidal and toroidal magnetic fields. As sample models, the
magnetized stars considered by Ioka and Sasaki~\cite{Ioka2004}, inside
which the magnetic fields are confined, are modified to the ones
stably stratified. The magnetized stars newly constructed in this
study are believed to be more stable than the existing relativistic
models because they have both poloidal and toroidal magnetic fields
with comparable strength, and magnetic buoyancy instabilities near the
surface of the star, which can be stabilized by the stratification,
are suppressed.
\end{abstract} 


\maketitle

\section{Introduction}

Recent observations established that soft-gamma repeaters (SGRs) and
anomalous x-ray pulsars (AXPs) are the so-called magnetars, i.e.,
highly magnetized neutron stars whose surface field strength is as
large as $\sim
10^{14}-10^{15}$~G~\cite{duncan,paczynsk,thompsona,thompsonb,woods}.
The presence of the magnetars has activated studies on equilibrium
configurations of magnetized stars, which have a long history.

Since the pioneering work by Chandrasekhar and Fermi \cite{chandra},
an enormous number of studies has been done for exploring structures
of magnetized stars: Prendergast~\cite{prendergast} and Woltjer
\cite{woltjer} calculated equilibrium configurations of
magnetized stars having mixed poloidal-toroidal fields, where the
magnetic fields are treated as first-order perturbations around a
spherical star (see, also, Ref.~\cite{roxburgh}).  Monaghan
\cite{monaghan} studied magnetized stars containing purely 
poloidal magnetic fields. Ioka~\cite{ioka} developed the works by
Prendergast and Woltjer into those of second-order perturbations to
study magnetic effects on the stellar structures. He also employed the
results obtained to explain magnetar activities.  Miketinac obtained
magnetized stars containing purely toroidal fields~\cite{miketinac}
and purely poloidal fields \cite{miketinac2} by solving exact master
equations numerically. Tomimura and Eriguchi developed a numerical
method for obtaining magnetized stars with mixed poloidal-toroidal
fields using a non-perturbative technique
\cite{tomimura} (see, also, Refs.~\cite{yoshida0,yoshida,lander}).
Duez and Mathis variationally considered the lowest-energy equilibrium states 
for a fixed magnetic helicity and constructed equilibria of magnetized stars having
mixed poloidal-toroidal fields by a perturbation technique~\cite{duez}.

General relativistic models of magnetized neutron stars have been also
explored.  Bocquet et al.~\cite{bocquet} and Cardall et
al.~\cite{cardall} obtained relativistic neutron star models with
purely poloidal magnetic fields. Using a perturbative technique, Konno
et al.~\cite{konno} calculated similar models. Kiuchi and Yoshida
\cite{kiuchi} computed magnetized stars with purely toroidal
fields.  Ioka and Sasaki~\cite{Ioka2004}, Colaiuda et
al.~\cite{colaiuda}, and Ciolfi et al.~\cite{ciolfia,ciolfib} derived
relativistic stellar models having both toroidal and poloidal magnetic
fields with a perturbative technique. Although progress has been
achieved in this field, further studies are required because all the
magnetized star models are constructed by some special magnetic-field
configurations which may not be realistic. In particular, it is not
clear at all whether their models are stable. 

The stability of the magnetized star is an important issue, because
only stable equilibrium models are viable.  Stability analyses of
magnetized stars have been performed by many works, since the
pioneering work by Tayler~\cite{tayler73}, who showed that stars
having purely toroidal magnetic fields are
unstable. Wright~\cite{wright73} subsequently showed that there is the
same type of the instability, the so-called pinch-type instability,
for stars containing purely poloidal magnetic fields. He also
suggested the possibility that stars having mixed poloidal-toroidal
magnetic fields may be stable if the strength of both components is
comparable (see, also, Refs.~\cite{markey,tayler80}).  Assche et
al.~\cite{assche} proved that the pinch-type instability in general
arises in magnetized stars with purely poloidal fields, and
Wright~\cite{wright73} and Markey and Tayler~\cite{markey} studied
this instability for particular magnetic-field configurations.
Flowers and Ruderman~\cite{flowers} found that another type of
instability occurs in purely poloidal magnetic-field configurations.

All those classical stability analyses have been done by a method of
an energy principle in the framework of Newtonian dynamics (see, also,
Refs.~\cite{pitts,goossens}).  Another approach is a local analysis,
with which Acheson \cite{acheson1978} investigated the stability of
rotating magnetized stars containing purely toroidal fields in detail
in the framework of Newtonian dynamics (see, also,
Refs.~\cite{acheson1979,spruit}) and derived detailed stability
conditions for purely toroidal magnetic fields buried inside rotating
stars with dissipation. Note that although it is an approximate
approach, the local analysis can take account of realistic effects on
the stability like rotation, heat conduction, and resistivity, which
cannot be included in a method of the energy principle.  Bonanno and
Urpin analyzed the axisymmetric stability~\cite{bonanno} and the 
non-axisymmetric stability~\cite{bonanno2} of cylindrical equilibrium
configurations possessing mixed poloidal-toroidal fields, while
ignoring compressibility and stratification of the fluid.

Recently the stability problem of the magnetized star has been
approached from another direction. By following the time evolution of
small random initial magnetic fields around a spherical star in the
framework of Newtonian resistive magnetohydrodynamics, Braithwaite and
Spruit~\cite{Braithwaite2004,Braithwaite2006} obtained stable
configurations of a magnetized star that are formed as a
self-organization phenomenon.  The resulting stable magnetic fields
have both poloidal and toroidal components with comparable strength
and support the conjecture for stability conditions of the magnetized
star given by the classical studies mentioned before.
By using the numerical magnetohydrodynamics simulation, Braithwaite
\cite{Braithwaite2009} studied stability conditions for the
magnetized stars and obtained a stability condition for his models
given in terms of the ratio of the poloidal magnetic energy to the
total magnetic energy which is of order unity.  Duez et al. showed
that magnetized stars constructed in Ref.~\cite{duez} exhibit no
instability for several Alfv\'{e}n time scales in their numerical
simulations \cite{duez10}. This fact reconfirms the results given by
Braithwaite.  Lander and Jones explored the stability of magnetized
stars by numerically solving the time evolution of linear
perturbations around the stars in their series of papers
\cite{lander11,lander11b,lander12}.  For the purely toroidal/poloidal
field cases, their results are consistent with those of the classical
stability analysis, i.e., the pinch-type instability is observed near
the symmetry and the magnetic axes for the purely toroidal and purely
poloidal field cases, respectively.  They also assessed the stability
of various magnetized stars with mixed poloidal-toroidal fields and
found that all their models considered suffer from the pinch-type
instability even for the cases in which the poloidal and toroidal
components have comparable strength \cite{lander12}. It is obvious
that the results by Lander and Jones are incompatible with those by
Braithwaite and his
collaborators~\cite{Braithwaite2009,duez10}. Lander and Jones
discussed the possibility that some physics missing in their study
would suppress the instability they found. We infer that in
particular, {\em stratification of the fluid} will be a key
ingredient, which is taken into account in the analyses of
Refs.~\cite{Braithwaite2009,duez10} but not in the analyses of
Ref.~\cite{lander12}. We will return to this point later.

General relativistic magnetized stars have been also analyzed
recently. By numerical-relativity simulations, Kiuchi et
al.~\cite{kiuchia,kiuchib} investigated the stability of the
magnetized stars with purely toroidal magnetic fields obtained by
Kiuchi and Yoshida~\cite{kiuchi}.  They showed that the stars with
some specific distributions of magnetic fields are stable against
axisymmetric perturbations but all the models considered are unstable
against non-axisymmetric perturbations due to the strong magnetic
buoyancy instability near the surface of the stars. The initial
behavior of the instability observed in Refs.~\cite{kiuchia,kiuchib}
is consistent with that expected by the Newtonian linear analyses by
Acheson~\cite{acheson1978}.  Lasky et al.~\cite{lasky} and Ciolfi et
al.~\cite{ciolfi} showed by numerical-relativity simulations that the
purely poloidal magnetic fields, obtained by Bocquet et
al.~\cite{bocquet}, are unstable due to the pinch-type instability
near the magnetic axis as predicted by the Newtonian linear
analyses~\cite{wright73,markey}. 

All these recent general relativistic magnetohydrodynamics simulations
have contributed a lot to the progress of the stability analyses of
general relativistic magnetized stars.  However, the numerical
simulations have been performed for equilibrium stars composed of {\em
  non-stratified} fluids. The assumption of non-stratification is
often used and quite reasonable as a first approximation for exploring
cold neutron stars, even though the cold neutron star is expected to
be highly stably stratified by the composition gradient (see, e.g.,
Refs.~\cite{finn,reisenegger,lai}).  If the effects of magnetic fields
are taken into account for the neutron star models in their stability
analysis, however, the situation changes drastically.  As discussed by
many authors, e.g., Reisenegger \cite{reisenegger2008} and Kiuchi et
al. \cite{kiuchib}, it has long been known that the magnetic buoyancy
makes the magnetized star unstable and that the stable stratification
is necessary to remove the magnetic buoyancy instability (the
so-called Parker instability~\cite{parker}).  It should be emphasized
that in the assumption that the stellar matter is stably stratified,
Braithwaite and Spruit~\cite{Braithwaite2004,Braithwaite2006} obtained
stable magnetized stars of simple magnetic configurations in their
numerical simulations.  Therefore, we infer that a stable
stratification is one of the key ingredients for stable configurations
of the magnetized stars. Note that Lander and Jones~\cite{lander12}
indicated the possibility that for stars containing mixed
poloidal-toroidal magnetic fields, some weak instability associated
with poloidal magnetic fields may not be removed by stable
stratification (see, also, Ref.~\cite{bonanno}). However, at the
moment, no definite conclusion has been obtained.  A reason that
non-stratified magnetized stars are employed for the stability
analyses in general relativity is that no model of stratified
magnetized stars have been constructed, although in the
framework of Newtonian dynamics, magnetized stars with stable
stratification due to composition gradients have been
obtained~\cite{mastrano,lander12b,glampedakis}. 

In this paper, thus, we study stably stratified and magnetized stars
in the framework of general relativity aiming at giving a prescription
for constructing them. First of all, we describe a general formulation
to obtain stationary axisymmetric magnetized stars composed of both
toroidal and poloidal magnetic fields with stratification due to
entropy gradients assuming ideal magnetohydrodynamics and employing a
barotropic equation of state. As sample models, the magnetized stars
considered by Ioka and Sasaki~\cite{Ioka2004}, which contain poloidal
and toroidal fields of comparable strength only inside the stars, are
modified to be the ones stably stratified.  Note that to date, no
general relativistic magnetized stars containing mixed
poloidal-toroidal fields have been constructed with a non-perturbative
approach because of difficulties in the treatment of non-circular
spacetimes (see, e.g., Ref.~\cite{erik}).  Finally, we describe the
reason that the magnetized stars obtained in the present study are
more stable than the existing relativistic models. 

\section{Basic equations for the general relativistic ideal magnetohydrodynamics}

We consider perfect fluids coupled with electromagnetic fields,
described by the basic equations summarized as follows.  The baryon
mass conservation equation:
\beq
\nabla_\mu(\rho u^\mu)=0\,,
\label{baryon_con}
\eeq
where $\rho$ and $u^\mu$ are the rest-mass density and the fluid four
velocity, respectively with $\nabla_\mu$ being the covariant
derivative associated with the metric $g_{\mu\nu}$.  The two sets of
the Maxwell equations:
\beqn
&&\nabla_\alpha F_{\mu\nu}+\nabla_\mu F_{\nu\alpha}+\nabla_\nu F_{\alpha\mu}=0\,, 
\label{maxwell1}
\\
&& \nabla_\nu F^{\mu\nu}=4\pi J^\mu\,, 
\label{maxwell2}
\eeqn
where $F_{\mu\nu}$ and $J^\mu$ are the Faraday tensor and the current
four vector, respectively.  The total energy-momentum conservation
law:
\beq
\nabla_\nu T^{\mu\nu}=0\,,
\label{devT}
\eeq
where $T^{\mu\nu}$ is the total energy-momentum tensor, defined by 
\beq
T^{\mu\nu}=\rho h u^\mu u^\nu + P g^{\mu\nu}+
{1\over 4\pi}\left[ F^{\mu\alpha}{F^\nu}_\alpha 
-{1\over 4}g^{\mu\nu}F^{\alpha\beta}F_{\alpha\beta}
\right]\,,
\eeq
with $h$ and $P$ being the specific enthalpy and the pressure,
respectively. Here, the specific enthalpy is written, in terms of the
specific internal energy $\varepsilon$, the pressure $P$, and the
rest-mass density $\rho$, by
\beq
h\equiv 1+\varepsilon+P/\rho\,. 
\label{Def_h}
\eeq

It is convenient to introduce the electric field $E_\mu$ and the
magnetic field $B_\mu$ observed by an observer associated with the
matter four velocity $u^\mu$, defined by
\beqn
E_\mu&=&F_{\mu\nu}u^\nu\,,\\
B_\mu&=&-{1\over 2}\epsilon_{\mu\nu\alpha\beta}u^\nu F^{\alpha\beta}\,,
\eeqn
where $\epsilon_{\mu\nu\alpha\beta}$ is the Levi-Civita tensor with
$\epsilon_{0123}=\sqrt{-g}$ with $g$ being the determinant of the
metric $g_{\mu\nu}$.  In a large number of intriguing astrophysical
problems, fluids coupled with electromagnetic fields can be
approximated with perfectly conductive ones. Thus, we further assume
the condition of perfect conductivity,
\beq
E_\mu=F_{\mu\nu}u^\nu=0\,.
\label{MHD_condition}
\eeq
The dual tensor of $F^{\mu\nu}$ is then
\beq
F^{*\mu\nu}=\frac{1}{2}\epsilon^{\mu\nu\alpha\beta}F_{\alpha\beta}
=B^\mu u^\nu-B^\nu u^\mu. \label{Dual}
\eeq

Equation~(\ref{devT}) is often decomposed into two sets of
equations, the energy equation and the momentum equation in the fluid
rest frame, respectively, given by
\beqn
-u_\mu \nabla_\nu T^{\mu\nu}&=&u^\nu
 \nabla_\nu\{\rho(1+\varepsilon)\}+\rho h \nabla_\nu u^\nu
\nonumber \\ 
&=&\rho u^\nu\nabla_\nu\varepsilon+P\nabla_\nu u^\nu=0\,,
\label{energy_eq}\\
q_{\mu\alpha}\nabla_\nu T^{\alpha\nu}&=&\rho h u^\nu\nabla_\nu
u_\mu+q_\mu^\nu\nabla_\nu P -F_{\mu\nu}J^\nu=0\,,
\label{momentum_eq}
\eeqn
where $q_{\mu\nu}\equiv g_{\mu\nu}+u_\mu u_\nu$. The perfect
conductivity condition has been used to derive Equations
(\ref{energy_eq}) and (\ref{momentum_eq}).  Using Equation
(\ref{baryon_con}) and the first law of thermodynamics,
\beq
d\varepsilon = {P\over \rho^2}d\rho+T dS \,, 
\label{1st_law}
\eeq
Equation (\ref{energy_eq}) is recasted into the entropy equation, 
\beq
u^\nu \nabla_\nu S=0 \,,
\label{entropy_eq}
\eeq
where $S$ and $T$ are the specific entropy and the temperature, respectively. 

To construct a magnetized star, we employ barotropic equations of
state given by
\beq
P=P(\rho)\,, \quad \varepsilon = \varepsilon(\rho)\,.
\label{eos}
\eeq
This equation of state is often used when studying a realistic star.

It should be noted that non-stratified isentropic fluids,
characterized by $d\varepsilon=\rho^{-2} Pd\rho$, have been assumed in
a large number of studies of neutron stars in general relativity. When
studying cold neutron stars in a chemical equilibrium, this
simplification may be accepted. However, this is not the case if
effects of magnetic fields are taken into account.  As pointed out by
many authors (see, e.g., Refs.~\cite{reisenegger2008} and
\cite{kiuchib}), the stable stratification is a necessary condition for
magnetized stars to be stable.  The reason is that magnetic flux tubes
inside the star basically suffer from magnetic buoyancy and are forced
to move toward their surface, resulting in a desperate instability of
the stars.  To counteract this magnetic buoyancy and to stabilize the
stars, the stable stratification is required. For sufficiently cold
neutron stars, the proton-neutron composition gradient is a candidate
for the stratification~\cite{reisenegger}. When studying on the
structure of a magnetized star, therefore, it is crucial to take into
account this effect.  Thus, the condition, $d\varepsilon=\rho^{-2}
Pd\rho$, which implies that the star is non-stratified, should not be
a priori assumed.  Note that whether the star is barotropic or not is
independent of whether the star is stably stratified or not. With the
one-parameter equation of state~(\ref{eos}), it is possible to have a
stably stratified star.

\section{Master equations for equilibrium magnetized stars with stable 
stratifications}

In this section, we derive the master equations for describing stably
stratified axisymmetric rotating stars composed of mixed
poloidal-toroidal magnetic fields.  We take the time and azimuthal
coordinates as $x^0=t$ and $x^3=\varphi$, respectively. Then,
components of the two Killing vectors may be written as
$t^\mu=\delta^\mu_0$ and $\varphi^\mu=\delta^\mu_3$, where
$\delta^\mu_\nu$ is the Kronecker delta. The other two spatial
coordinate variables are written as $x^1$ and $x^2$ in this section.
Thus, the quantities describing the equilibrium stars are basically
functions of $x^1$ and $x^2$ only.  Henceforth, capital Latin indices
($A, B, C, \cdots$) run from 1 to 2.

Because of the assumption of the axial symmetry and stationarity,
Equation~(\ref{entropy_eq}) is written as
\beq
u^A\partial_A S=0\,. \label{circ}
\eeq
This equation means that the specific entropy has to be constant along
streamlines on the stellar meridional plane unless $u^A=0$. However,
the constant specific entropy distribution along streamlines is not
realized for stable magnetized stars because of the following reason:
For stationary axisymmetric stars, streamlines on the meridional
plane, in general, are closed curves.  Thus, it is inevitable that
there exits an unstably stratified region (convectively unstable
region) where $(\nabla^\mu P)(\nabla_\mu S) < 0$ is satisfied.  To
construct a stably stratified (convectively stable) star, thus, we
have to assume
\beq
u^A=0. \label{no_meridional_flow}
\eeq
Then, we have no condition for $S$ apart from the assumption of the
stationarity and the axial symmetry.  In other words, we can freely
choose a functional form of $S$ if $u^A=0$ is assumed. 

The assumption, $u^A=0$, is an essential difference between our study
and the study of Ref.~\cite{Ioka2004} in which the {\it isentropic} meridional flow 
is taken into account.  The fluid four-velocity is, thus,
given by
\beq
u^\mu=\gamma (t^\mu+\Omega \varphi^\mu)\,. 
\label{Def_u}
\eeq
where $\Omega$ is the angular velocity of the fluid and $\gamma=u^0$.

From Equations (\ref{maxwell1}), (\ref{MHD_condition}), and the
integrability condition for the momentum equation~(\ref{momentum_eq}),
we have the following relations:
\beq
F_{A3}={\partial\over\partial x^A} \Psi\,,
\label{Def_Psi}
\eeq
\beq
F_{03}=0\,,
\eeq
\beq
F_{0A}=-\Omega_F(\Psi) F_{3A}\,,
\label{Def_omgF}
\eeq
\beq
\sqrt{-g} F^{12}=\hat{\Gamma}(\Psi)\,,
\label{Def_hatGamma}
\eeq
\beq
-\Omega_F(\Psi) u^0+u^3=0\,, 
\label{constraint_omgF}
\eeq
\beq
-\ln\gamma+\int{dP\over\rho h}-\int\mu(\Psi)d\Psi+\hat{C}=0\,,
\label{Def_Lambda}
\eeq
\beq
J^3-\Omega_F J^0=\rho h\{\mu(\Psi)+\gamma u_3\Omega_F'
\}+{F_{12}\over4\pi\sqrt{-g}}\hat{\Gamma}'\,,
\label{current}
\eeq
where $\Omega_F$, $\hat{\Gamma}$, and $\mu$ are arbitrary functions of
the flux function $\Psi$, which is the azimuthal component of the
vector potential $A_\mu$ associated with $F_{\mu\nu}$ as shown in
Equation (\ref{Def_Psi}). Here, $\hat{C}$ is an integral constant, and the
prime denotes the derivative with respect to $\Psi$. Note that
$\Omega_F$ is sometimes called the angular velocity of the magnetic
field line. Substituting Equation (\ref{Def_u}) into Equation
(\ref{constraint_omgF}), we obtain
\beq
\Omega_F=\Omega\,. 
\eeq
Equation (\ref{Def_Lambda}) corresponds to the equation of the
hydrostatic equilibrium.

For cases of non-stratified stars,  
the present formulation may be derived from that of Ioka and
Sasaki~\cite{Ioka2004} by taking a limit.  First, we focus on
Equations (16) -- (18) of Ref.~\cite{Ioka2004}. Using a relation
\beq
\gamma (u_0+\Omega u_3) = -1\,,
\eeq
or $u^\mu u_\mu=-1$, we may rewrite Equation (\ref{Def_Lambda}) as 
\beq
(u_0+\Omega u_3)\,e^{\displaystyle \int{dP\over\rho h}} = -
e^{\displaystyle \int\mu(\Psi)d\Psi-\hat{C}}\,,
\eeq
which corresponds to Equation (16) of Ref.~\cite{Ioka2004}.  We
therefore found that functions $\mu$ and $D$ in Ref.~\cite{Ioka2004}
correspond to $\exp(\int{dP\over\rho h})$ and
$\exp(\int\mu(\Psi)d\Psi-\hat{C})$, respectively.
Note that $\mu$ of Ref.~\cite{Ioka2004} is exactly the same as $h$ of
the present study and that $h^{-1} dh = (\rho h)^{-1} dP$ in the
isentropic case $d\varepsilon=\rho^{-2} Pd\rho$.  Equations (17) and
(18) of Ref.~\cite{Ioka2004} are, respectively, derived from the
energy and angular momentum conservation equations,
\beq
\nabla_\mu ({T^\mu}_\nu t^\nu) =0\,, \quad \nabla_\mu ({T^\mu}_\nu
\varphi^\nu) =0\, .
\eeq
These may be explicitly written by
\beq
\nabla_\mu \left(\rho h u_\nu \xi^\nu u^\mu\right)+
\nabla_\mu \left[{1\over 4\pi} F^{\mu\alpha} F_{\nu\alpha}
\xi^\nu\right]  =0\, , 
\label{con_eq}
\eeq
with $\xi^\mu$ being $t^\mu$ or $\varphi^\mu$. Since we assume the
condition (\ref{no_meridional_flow}), the first term in the left-hand
side of Equation (\ref{con_eq}) automatically vanishes and we obtain
two relations,
\beq
X = \sqrt{-g} F^{12} \Omega \,, \quad Y = \sqrt{-g} F^{12}  \, , 
\label{Def_XY}
\eeq
where $X$ and $Y$ are arbitrary functions of the flux function
$\Psi$. These relations were already derived in Equations
(\ref{Def_omgF}) and (\ref{Def_hatGamma}).  Thus, we find that
$X=\hat{\Gamma}\Omega$ and $Y=\hat{\Gamma}$. The functions $X$ and $Y$
are, in terms of the functions of Ref.~\cite{Ioka2004}, $C$, $E$, and
$L$, given by
\beq
X =-4\pi {E\over C} \,, \quad Y = -4\pi{L \over C} \, . 
\eeq
From Equation (20) of Ref.~\cite{Ioka2004}, we have $D/C=E/C -\Omega
L/C$. As argued in Ref.~\cite{Ioka2004}, the no-meridional flow limit
of magnetized stars with mixed poloidal-toroidal magnetic fields is
given by the limit $C\rightarrow \infty$ with $D$, $L/C$ and $E/C$
kept to be finite.  Thus, we find that $D/C=E/C -\Omega L/C$ becomes
$0=-4\pi E/C +4\pi \Omega L/C=X-\Omega Y$ in this limit, which is
automatically satisfied by Equation (\ref{Def_XY}) in the present
situation. 

Once the metric coefficients are given, all the components of
$F^{\mu\nu}$ are written as functions of $\Psi$ and
$\partial\Psi/\partial x^A$ through the two arbitrary functions of the
flux function, $\Omega_F$ and $\hat{\Gamma}$.  Using the Maxwell
equation~(\ref{maxwell2}), thus, $J^0-\Omega_F J^3$ appearing in the
left-hand side of Equation~(\ref{current}) can be written as a
second-order elliptic-type partial differential operator for
$\Psi(x^1, x^2)$. Equation~(\ref{maxwell2}) in conjunction with
Equation~(\ref{current}) may then be solved to obtain a distribution
of the magnetic fields around a star. This equation is often
called the Grad-Shafranov (GS) equation~\cite{Lovelace} when it is
written as a partial differential equation for $\Psi$. 

It is useful to introduce some global quantities to characterize
equilibrium solutions of stars.  For equilibrium states of magnetized
stars, the total baryon rest mass $\widetilde{M}^*$, the internal
thermal energy $\widetilde{E}_{\rm int}$, and the electromagnetic
energy $\widetilde{E}_{\rm EM}$ may be defined as
\beqn
&&\widetilde{M}^*=\int\rho\gamma\sqrt{-g}\,d^3x\,, \label{baryon}\\
&&\widetilde{E}_{\rm int}=\int\rho\varepsilon\gamma\sqrt{-g}\,d^3x\,,\\
&&\widetilde{E}_{\rm EM}={1\over 8\pi}\int B^\mu B_\mu \gamma \sqrt{-g}\,d^3x\ \label{magnetic}\,.
\eeqn
(see, e.g., Ref.~\cite{kiuchi}.)

\section{Magnetic fields around a spherical star and their effects on
the stellar structures}

\subsection{Spherical stars with no magnetic field}

We assume that the magnetic energy density is much smaller than the
matter density and pressure so that the magnetic-field effects can be
treated as perturbations on a spherical non-magnetized star.  The
background metric is then given by
\beqn
ds^2=-e^{2\nu}dt^2+e^{2\lambda}dr^2+r^2(d\theta^2+\sin^2\theta d\varphi^2)\,, 
\eeqn
where $\nu$ and $\lambda$ are functions of $r$~\cite{mtw}. The function $\gamma$ 
for the spherical stars is written as  
\beq
\gamma=e^{-\nu}\,.
\eeq
The equilibrium state of a star is described by the set of the
following TOV equations~\cite{mtw}:
\beqn
&&{dm\over dr}=4\pi r^2 \rho(1+\varepsilon)\,,\label{eq:enclose}\\
&&{dP\over dr}=-e^{2\lambda}\rho h{m+4\pi P r^3 \over r^2}\,,\\
&&{d\nu\over dr}=-{1 \over \rho h}{dP\over dr}\,, 
\eeqn
where $m$ is defined by 
\beq
m\equiv {r \over 2}(1-e^{-2\lambda})\,. 
\eeq

For the unperturbed spherical stars, the gravitational mass $M$, the
total baryon rest mass $M^*$, and the internal thermal energy $E_{\rm
int}$ are given by
\beqn
&&M=m(R)\,, \\
&&M^*=4\pi \int_0^R \rho e^{\lambda} r^2 dr \,, \\ 
&&E_{\rm int}=4\pi \int_0^R \rho\varepsilon e^{\lambda} r^2 dr\,, 
\eeqn
where $R$ denotes the circumferential radius of the star defined by
$P(R)=0$.  The gravitational potential energy $W$ for the unperturbed
stars is defined by
\beqn
|W|=M^*+E_{\rm int}-M\,. 
\eeqn

\subsection{Magnetic field around a spherical star}

A profile of magnetic fields is determined by specifying the
functional forms of $\Omega_{\rm F}$, $\hat{\Gamma}$, and $\mu$, which
are the arbitrary functions of $\Psi$.  Following Ioka and
Sasaki~\cite{Ioka2004}, we assume that these three functions as well
as the integral constant, $\hat{C}$, are given by
\beqn
&&\Omega_F=\Omega=\Omega_2 \Psi^2\,, 
\label{Def_f_Om}\\
&&\mu=C_1 \,, 
\label{Def_f_mu}\\
&&\hat{\Gamma}=L \Psi \,, 
\label{Def_f_Gam} \\
&&\hat{C}=C_0+ C_2 \,, 
\label{Def_f_C}
\eeqn
where $\Omega_2$, $C_0$, $C_1$, $C_2$, and $L$ are constants. 
Because we consider weak magnetic fields around a spherical star, 
it is useful to introduce a smallness parameter $\eta$ for which $\Psi=O(\eta)$. 
For the constants appearing in Equations (\ref{Def_f_Om}) -- (\ref{Def_f_C}), 
we further assume that 
\beqn
&&\Omega_2 =O(1)\,, \  C_0=O(1)\,,\  C_1=O(\eta)\,, \nonumber \\ 
&& C_2=O(\eta^2)\,, L=O(1) \,, 
\eeqn
$C_0$ is determined in the background equation for Equation~(\ref{Def_Lambda}). 
Then, up to the first order in $\eta$, $F_{\mu\nu}$ is written as
\beq
F_{\mu\nu}=\left(
\begin{array}{cccc}
0 & 0 & 0& 0 \\
0 & 0 & e^{\lambda-\nu}L\csc\theta\Psi & \partial_r \Psi \\
0 & -e^{\lambda-\nu}L\csc\theta\Psi & 0 & \partial_\theta \Psi \\
0 & -\partial_r \Psi &-\partial_\theta \Psi & 0
\end{array}
\right)\,,
\eeq
and Equation (\ref{current}) becomes 
\beq
J^3=\rho h C_1+{e^{-2\nu}L^2\Psi\over 4\pi r^2\sin^2\theta}\,. 
\label{current2}
\eeq
Note that the contribution of $\Omega_2$ is neglected because it is a
higher-order quantity.  In terms of $\Psi$, $J^3$ is given by
\beqn
J^3&=&{1\over 4\pi}\nabla_\alpha F^{3\alpha} \nonumber \\ &=&-{1\over
4\pi e^{\nu+\lambda} r^2\sin^2\theta}
\left[
{\partial\over\partial r}\left(e^{\nu-\lambda}{\partial\over\partial
r}\Psi\right)\right. \nonumber \\
&&\vspace{3cm} \left.+{e^{\nu+\lambda}\over
r^2}\sin\theta{\partial\over\partial\theta}\left(
{1\over\sin\theta}{\partial\over\partial\theta}\Psi \right)
\right]\,. \label{current3}
\eeqn
From Equations~(\ref{current2}) and (\ref{current3}), we obtain the
master equation for the flux function $\Psi$ (the GS equation),
\beqn
&&e^{\lambda-\nu}
\left[
{\partial\over\partial r}\left(e^{\nu-\lambda}{\partial\over\partial
r}\Psi\right)+{e^{\nu+\lambda}\over
r^2}\sin\theta{\partial\over\partial\theta}\left(
{1\over\sin\theta}{\partial\over\partial\theta}\Psi \right)
\right]
\nonumber \\
&&+4\pi r^2\sin^2\theta \rho h e^{2\lambda} C_1+e^{2(\lambda-\nu)}L^2\Psi=0\,. 
\label{GS_eq}
\eeqn
Because it is the azimuthal component of the vector potential, the
flux function $\Psi$ may be expanded by the vector harmonics with the
axial parity as
\beq
\Psi=4\pi C_1 \sum_{l=1}^\infty r^{l+1} \psi_l(r)\sin\theta
{\partial\over \partial\theta}P_l(\theta)\,, 
\label{expand_Psi}
\eeq
where $P_l$ is the Legendre polynomial (see, e.g., Ref.~\cite{thorne}). Substituting Equation
(\ref{expand_Psi}) into the GS equation (\ref{GS_eq}) yields
\beqn
&&{d^2\psi_l\over dr^2}+\biggl({d(\nu-\lambda)\over dr}+{2(l+1)\over
r}\biggr){d\psi_l\over dr}
\nonumber \\
&&+\left[e^{2(\lambda-\nu)}L^2+{l(l+1)\over r^2}(1-e^{2\lambda})
+{d(\nu-\lambda)\over dr}{l+1\over r}\right]\psi_l
\nonumber \\
&&-\rho h \,e^{2\lambda} \delta^1_l=0\,. 
\label{GS2}
\eeqn

Regular solutions of Equation~(\ref{GS2}) near the center of the star
can be written as
\beqn
\psi_l=a_0+a_2 r^2+\cdots \,, \label{psi_bs1}
\eeqn
where $a_0$ and $a_2$ are constants with $a_2$ given by 
\beqn
a_2=&&-\frac{e^{-2 \nu_0}L^2-2(l+1)[(l+1)\lambda_2-\nu_2]}{2(2l+3)}\,a_0
\nonumber \\
&&+\frac{1}{10} \rho_0 h_0\,\delta^1_l\,. 
\eeqn
Here, constants $\nu_0$, $\nu_2$, $\lambda_2$, $\rho_0$ and $h_0$ are
defined in the power series expansion of the background quantities
near $r=0$ as follows:
\beqn
&&\nu=\nu_0+\nu_2 r^2+\cdots\,, \\
&&\lambda=\lambda_2 r^2+\cdots\,,\\
&&h=h_0+h_2 r^2+\cdots\,,\\
&&\rho=\rho_0+\rho_2 r^2+\cdots \,.
\eeqn

Following Ref.~\cite{Ioka2004}, we focus on magnetized stars whose
exterior is vacuum and whose surface has no magnetic field.  At the
surface of the star, then, we need to require two boundary conditions
for the flux function, given by
\beqn
\psi_l=0\,, \ 
{d\psi_l\over dr}=0\,, \quad {\rm at}\ r=R\,,
\label{surface_bc}
\eeqn
where $R$ is the radius of the unperturbed star. For the $l\ne 1$
cases, Equation (\ref{GS2}) becomes a homogeneous equation. In
general, then, the three boundary conditions, the regularity condition
at the center of the star and the surface boundary conditions given by
Equation (\ref{surface_bc}), cannot be satisfied simultaneously.  In
other words, we have to require $\psi_l=0$ for $l\ne 1$.  For the
$l=1$ case, due to the two boundary conditions at the stellar surface,
given in Equation (\ref{surface_bc}), the GS equation becomes an
eigenvalue equation with respect to the two parameters $a_0$ and $L$.
The other parameter $C_1$ can be assigned freely and determines the
strength of the magnetic fields. The remaining constant $C_2$ is
related to the pressure perturbation as discussed below.

The $r$ and $\theta$ components of the vector potential $A_\mu$ may be
obtained straightforwardly.  If we set $A_\theta=0$ by using a gauge
degree of freedom, $F_{12}$ is given by
\beqn
F_{12}&=&-\partial_\theta A_r \nonumber \\ 
&=&e^{\lambda-\nu}L\csc\theta\Psi \nonumber \\
&=&4\pi C_1\, e^{\lambda-\nu}L r^{2} \psi_1(r){\partial\over
\partial\theta}P_1(\theta)\,.
\eeqn
Requiring the regularity on the symmetry axis, we have 
\beqn
A_r=-4\pi C_1\, e^{\lambda-\nu}L r^{2} \psi_1(r) P_1(\theta)\,.
\eeqn
If another gauge condition is required, we may make a gauge transformation 
\beqn
A_\mu \rightarrow A_\mu-\partial_\mu \{f(r)P_1(\theta)\}\, ,
\eeqn
to obtain the vector potential that satisfies a required gauge condition.

\subsection{Stellar deformation due to the magnetic fields}

As discussed before, the flux function in the present situation is given by 
\beqn
\Psi&=&4\pi C_1\,  r^2 \psi_1(r)\sin\theta{\partial\over
\partial\theta}P_1(\theta)\,\nonumber \\
&\equiv&4\pi C_1\,  \Psi_1(r)\left({2\over 3}P_2(\theta)-{2\over 3}\right)
\,. 
\label{expand_Psi_l1}
\eeqn
Thus, the energy-momentum tensor associated with the electromagnetic
fields ${T_{\rm (em)}}^\mu_\nu$, defined by 
\beqn
&&{T_{\rm (em)}}^\mu_\nu=
{1\over 4\pi}\left[ F^{\mu\alpha}{F}_{\nu\alpha} 
-{1\over 4}\delta^{\mu}_{\nu}F^{\alpha\beta}F_{\alpha\beta}
\right]\,, 
\eeqn
induces a deviation of the order of $O(\eta^2)$ from the background spherical matter distribution. 
The line element is then perturbed as follows: 
\beqn
ds^2
 &=&-e^{2\nu}\left\{1+2 \sum_{i=0,2} (4\pi
 C_1)^2H_{i}(r)P_{i}(\theta)\right\}
dt^2 \nonumber \\
 &+&e^{2\lambda}\left\{1+2 {e^{2\lambda}\over r}\sum_{i=0,2} (4\pi
 C_1)^2M_{i}(r)P_{i}(\theta)\right\}dr^2 \nonumber \\
 &+&r^2\left\{1+2 (4\pi C_1)^2 K_{2}(r)P_{2}(\theta)\right\}
(d\theta^2+\sin^2\theta d\varphi^2) \nonumber \\
 &+&2 (4\pi C_1)^2W_{2}(r)\sin\theta\partial_\theta P_{2}(\theta)drd\varphi \,,
\label{perturbed_metric}
\eeqn
where $H_i$, $M_i$, $K_i$, $I_i$, $V_i$, $W_i=O(1)$ because
$C_1=O(\eta)$. Here, we employ the Regge-Wheeler gauge. In this
perturbed spacetime, the function $\gamma$ in Equation (\ref{Def_u})
is given by
\beqn
\gamma=e^{-\nu}\left\{1-\sum_{i=0,2} (4\pi C_1)^2H_{i}(r)P_{i}(\theta)\right\}\, 
\eeqn
because $\delta u^A=0$ and $\delta u^3$ is the second-order quantity
(see Equations~(\ref{no_meridional_flow}) and (\ref{Def_f_Om})).
Here, we have omitted the terms higher than $O(\eta^2)$. From Equation
(\ref{Def_Lambda}), the pressure perturbation $\delta P$ is, in terms
of the metric and the flux functions, written as
\beqn
\delta P&\equiv&(4\pi C_1)^2\sum_{i=0,2} \delta P_{i}(r)P_{i}(\theta)  \nonumber \\
&=&\rho h \left\{ 
C_1\Psi-C_2-\sum_{i=0,2} (4\pi C_1)^2H_{i}(r)P_{i}(\theta) \right\}.
\nonumber \\ 
\eeqn
Thus, we have  
\beqn
\delta P_0&=&\rho h \left[-{1\over 6\pi}\Psi_1(r)-H_{0}(r)-\tilde{C}_2 \right] \,,  \label{dp0eq} \\
\delta P_2&=&\rho h \left[ {1\over 6\pi}\Psi_1(r)-H_{2}(r)\right] \,. \label{dp2eq}
\eeqn
where $\tilde{C}_2\equiv C_2/(4\pi C_1)^2$. 

We then obtain a set of the metric perturbation equations as follows:
\beqn
{dM_0\over dr}&=&4\pi r^2 \rho h \left( {d(\rho+\rho\varepsilon)\over d P}\right) \left({\delta P_0\over \rho h}\right)
+{1\over 3}e^{-2\lambda}\left({d\Psi_1\over dr}\right)^2  \nonumber \\
&&+{1\over 3r^2}\left(2+L^2 r^2 e^{-2\nu}\right)({\Psi_1})^2 \,, 
\label{m0}
\eeqn
\beqn
{dH_0\over dr}&=&4\pi r e^{2\lambda} \rho h  \left({\delta P_0\over \rho h}\right) \nonumber \\ 
&&+{e^{2\lambda}\over r^2}\left(1+2r{d\nu\over dr}\right)M_0 
+{1\over 3r}\left({d\Psi_1\over dr}\right)^2 \nonumber \\  
&&+{e^{2\lambda}\over 3 r^3}\left(-2+L^2 r^2 e^{-2\nu}\right)({\Psi_1})^2\,,
\label{h0}
\eeqn
\beqn
W_2={2\over 3}L e^{\lambda-\nu}({\Psi_1})^2 \,, \label{w2eq}
\eeqn
\beqn
H_2+{e^{2\lambda} M_2 \over r}={2e^{-2\lambda}\over 3}\left({d\Psi_1\over dr}\right)^2-{2e^{-2\nu}\over 3} L^2 ({\Psi_1})^2 \,, 
\label{m2eq} 
\eeqn
\beqn
&&{1\over r}\left({dH_2\over dr}+{dK_2\over dr}\right)+{d\nu\over dr}{dK_2\over dr}
-2{e^{2\lambda}\over r^2}(K_2+H_2) \nonumber \\
&&-{e^{2\lambda}\over r^2}H_2-{e^{2\lambda}\over r^3}\left(1+2r{d\nu\over dr}\right)M_2 \nonumber \\
&&=4\pi e^{2\lambda} \delta P_2 -{1\over 3r^2}\left({d\Psi_1\over dr}\right)^2 \nonumber \\
&&-{e^{2\lambda}\over 3r^4}(4+L^2 r^2 e^{-2\nu})({\Psi_1})^2 \,, 
\eeqn
\beqn
&&{dH_2\over dr}+{dK_2\over dr}+{1\over r}\left(-1+r{d\nu\over dr}\right)H_2  \nonumber \\
&&\quad -{e^{2\lambda}\over r^2}\left(1+r{d\nu\over dr}\right)M_2 
={4\over 3r^2}{\Psi_1}{d\Psi_1\over dr} \,.\label{h2}
\eeqn

For the perturbation with $l=0$, it is convenient to obtain $M_0$ and
$\delta P_0/ (\rho h)$ first. The equation for determining $\delta
P_0/ (\rho h)$ is derived from Equations~(\ref{dp0eq}) and (\ref{h0})
as
\beqn
{d\over dr}\left({\delta P_0\over \rho h}\right)&=&-{1\over
6\pi}{d\Psi_1\over dr}-{dH_0\over dr} \nonumber \\ &=&-{1\over
6\pi}{d\Psi_1\over dr}-4\pi r e^{2\lambda}\rho h \left({\delta
P_0\over \rho h}\right) \nonumber \\ &&-{e^{2\lambda}\over
r^2}\left(1+2r{d\nu\over dr}\right)M_0 -{1\over 3r}\left({d\Psi_1\over
dr}\right)^2 \nonumber \\ &&-{e^{2\lambda}\over 3 r^3}\left(-2+L^2 r^2
e^{-2\nu}\right)({\Psi_1})^2\,.
\eeqn
Following Ref.~\cite{Ioka2004}, a new dependent variable $Y_2$ is, for
the perturbation with $l=2$, introduced by
\beqn
Y_2&\equiv&H_2+K_2  \label{def_Y}  \\  
&-&{e^{-2\lambda}\over 6}\left\{ \left({d\Psi_1\over dr}\right)^2+{4\over r}\Psi_1{d\Psi_1\over dr}+
{4 e^{2\lambda}\over r^2}(\Psi_1)^2 \right\} \,, \nonumber 
\eeqn
which facilitates the numerical computation. Then, two independent
variables $Y_2$ and $H_2$ may be determined by
\beqn
{dY_2\over dr}&=&-2{d\nu\over dr}H_2-{r\rho h \over 3}\left(2\Psi_1+r {d\Psi_1\over dr}\right) 
\nonumber \\
&-& {2\over 3}e^{-2\nu}L^2{d\nu\over dr}(\Psi_1)^2 +e^{-2\lambda}{d\nu\over dr}\left({d\Psi_1\over dr}\right)^2  \\ 
&+& {1\over 3r^2}
\left[ -2+ 2e^{-2\lambda}\left\{1+r{d(\nu+\lambda)\over dr}\right\} \right. \nonumber \\ 
&& \left. \quad\quad\quad\quad +e^{-2\nu}L^2 r^2 \right]\Psi_1 {d\Psi_1\over dr} \nonumber \,, 
\eeqn
\beqn
{d\nu\over dr}{dH_2\over dr}&=&\left\{{1\over r^2}(1-e^{2\lambda})-2\left({d\nu\over dr}\right)^2 +4\pi e^{2\lambda}h\rho \right\}H_2
 \nonumber \\
&-&2{e^{2\lambda}\over r^2}Y_2 -{2\over 3} e^{2\lambda}\rho h\,\Psi_1  \\
&+&{2\over 3}e^{-2\lambda}\left({d\nu\over dr}\right)^2\left({d\Psi_1\over dr}\right)^2
+{4\over 3 r^2}{d\nu\over dr} \,\Psi_1{d\Psi_1\over dr} \nonumber \\
&+&{e^{-2\nu}L^2\over 3 r^2}\left\{e^{2\lambda}-2 r^2\left({d\nu\over dr}\right)^2\right\}(\Psi_1)^2 \nonumber \,. 
\eeqn
Once $\Psi_1$, $\delta P_0/(\rho h)$, $M_0$, $Y_2$, and $H_2$ are obtained, the other perturbation quantities may be calculated 
algebraically through Equations (\ref{dp0eq}), (\ref{dp2eq}), (\ref{w2eq}), (\ref{m2eq}), and (\ref{def_Y}). 

Near the center of the star, the physically acceptable solutions may be expanded in the power series of $r$ as 
\beqn
\delta P_0/(\rho h)&=&h_{00}+h_{02} r^2+\cdots \,, \label{exp1} \\
M_0&=& r^3(m_{00}+m_{02} r^2\cdots) \,, \\
Y_2&=& r^4(y_{20}+y_{22} r^2+\cdots) \,, \\
H_2&=& r^2(h_{20}+h_{22} r^2+\cdots)\,. \label{exp4} \
\eeqn
Here, we have two options for determining a value of $h_{00}$. One is
to set $h_{00}=0$, which corresponds to considering sequences of the
magnetized stars characterized by the fixed central density. The other
is to use $h_{00}$ to keep the total baryon mass $\widetilde{M}^*$
constant for the magnetized stars.  This corresponds to considering
sequences of the magnetized stars characterized by the fixed total
baryon mass. Following Ref.~\cite{Ioka2004}, we choose the latter
option, i.e., on the constant baryon mass sequences of the magnetized
stars.

Outside the star, the master equations become
\beqn
{dM_0\over dr}&=&0\,, 
\eeqn
\beqn
{dH_0\over dr}
={M_0\over (r-2M)^2}\,,
\eeqn
\beqn
W_2=0 \,, 
\eeqn
\beqn
H_2+{M_2 \over r-2 M}=0 \,, 
\eeqn
\beqn
{dY_2\over dr}=-{2M\over r(r-2M)}H_2 \,, \label{oeq1}
\eeqn
\beqn
{dH_2\over dr}&=&-2\left\{{1\over r}+{M\over r(r-2M)} \right\}H_2 -{2\over M}Y_2 \,. \label{oeq2}
\eeqn
For the $l=0$ perturbation, the vacuum solutions are given by  
\beqn
M_0=M_0(R)={\rm const.} \,, \quad H_0=-{M_0(R)\over r-2M}\,. \label{out_sol0}
\eeqn
As for the $l=2$ perturbation, manipulating Equations (\ref{oeq1}) and
(\ref{oeq2}) yields
\beqn
&&-(y+1)(y-1){d^2H_2\over dy^2}-2y{dH_2\over dy}  \\
&&\quad\quad+\left\{6+{4 \over (y+1)(y-1)} \right\}H_2=0 \,, \nonumber
\eeqn
where $y\equiv r/M-1$. This is the associated Legendre equation. Since
$\displaystyle{\lim_{y \rightarrow \infty} H_2=0}$ for physically
acceptable solutions, we obtain
\beqn
H_2=D Q_2^2(y)\,, \label{out_sol1}
\eeqn
where $Q_l^m$ and $D$ are the associated Legendre function of the
second kind and a constant, respectively.  With a recurrence relation
for $Q_l^m$,
\beqn
{dQ^m_l\over dy}={(l+m)(l-m+1)\over \sqrt{y^2-1}}Q^{m-1}_l-{my\over y^2-1}Q^m_l \,. 
\eeqn
we have 
\beqn
Y_2&=&-{2 D\over \sqrt{y^2-1}}Q_2^1(y) \,. \label{out_sol2}
\eeqn
At the surface of the star, the outer solutions, given by
(\ref{out_sol0}), (\ref{out_sol1}) and (\ref{out_sol2}), are matched
to the inner solutions integrated from the center of the star with the
boundary conditions (\ref{exp1}) -- (\ref{exp4}).

\subsection{Global quantities characterizing magnetized stars}

As mentioned before, the global physical quantities~(\ref{baryon}) --
(\ref{magnetic}) are useful for exploration of the magnetized
star. Perturbations due to the magnetic effects in the gravitational
mass, the total baryon rest mass $\widetilde{M}^*$, and the internal
thermal energy $\widetilde{E}_{\rm int}$ are, respectively, given by
\beqn
&&\Delta M = (4\pi C_1)^2 M_0(R) \,, \\ 
&&\Delta M^*=4\pi (4\pi C_1)^2
\nonumber \\ 
&&\quad\quad \times \int_0^R \rho e^{\lambda} r^2
\left({d\ln\rho\over dP}\delta P_0+{e^{2\lambda}M_0\over r}\right) dr
\,, \\ 
&&\Delta E_{\rm int}=4\pi (4\pi C_1)^2 \nonumber \\ &&\quad
\times \int_0^R \rho\varepsilon e^{\lambda} r^2
\left({d\ln(\rho\varepsilon)\over dP}\delta P_0+{e^{2\lambda}M_0\over
r} \right)dr\,.
\eeqn
As already mentioned, we study the sequences of equilibrium states of
the magnetized star characterized by the fixed total baryon
mass. Thus, the condition of $\Delta M^*=0$ is employed for
determining values of $h_{00}$ in Equation (\ref{exp1}), which is
related to the perturbations in the central density of the star,
$\Delta\rho_c$, through the relation
\beqn
\Delta\rho_c=(4\pi C_1)^2 \rho h {d\rho\over dP}\Big|_{r=0}\, h_{00}  \,. 
\eeqn
The electromagnetic energy $E_{\rm EM}$ is decomposed as 
\beqn
E_{\rm EM}=E_{\rm EM}^{(p)}+E_{\rm EM}^{(t)} \,, 
\eeqn
where $E^{(p)}_{\rm EM}$ and $E_{\rm EM}^{(t)}$ are the poloidal and
toroidal magnetic-field energies, respectively, given by
\beqn
&&E_{\rm EM}^{(p)}={16\pi^2 C_1^2\over 3} \nonumber \\ 
&& \times
\int_0^R \left[ e^{-\lambda}\left\{{d\over
dr}(r^2\psi_1)\right\}^2+2e^{\lambda} r^2\psi_1^2 \right] \,dr\ \,, \\
&&E_{\rm EM}^{(t)}={16\pi^2 C_1^2\over 3}\int_0^R L^2 e^{\lambda-2\nu}
r^4\psi_1^2 \,dr \,.
\eeqn
Multipole moments are also global and physical quantities
characterizing the equilibrium star.  The constant $D$ appearing in
the outer solutions, Equations (\ref{out_sol1}) and (\ref{out_sol2}),
is related to the mass quadrupole moment $\Delta Q$, defined by
$\Delta Q \equiv 8 M^3 D/5$ (see, e.g., Refs.~\cite{Ioka2004,thorne}).

Characteristic quantities for the stellar deformation due to magnetic
stress also feature the magnetized stars. The surface of the star is
defined by $P(r)+(4\pi C_1)^2(\delta P_0(r)+\delta P_2(r)
P_2(\theta))=0$. Thus, the radial displacement of the fluid elements
on the stellar surface, $\Delta r$, is given by
\beqn
\Delta r&=&(\Delta r)_0+(\Delta r)_2 P_2(\theta) \,, \nonumber \\
&=&-(4\pi C_1)^2 (\delta P_0(R)+\delta P_2(R) P_2(\theta)){dr \over dP}(R) \,. 
\eeqn
Here, $(\Delta r)_0$ may be interpreted as an average change in the
radius of the star induced by the magnetic effects.  The degree of the
quadrupole surface deformation due to the magnetic stress is well
described by the ellipticity, given by
\beqn
e^*=-{3\over 2}(4\pi C_1)^2 \left[ {(\Delta r)_2 \over R}+K_2(R) \right] \,, 
\eeqn
where $e^*$ is defined as a relative difference between the equatorial
and polar circumference radii of the star \cite{Ioka2004}.  Thus,
$e^* <0$ ($e^* >0$) means that the star is prolate (oblate).

Another physically important quantity of magnetized objects is the
total magnetic helicity ${\cal H}$, which is conserved in ideal
magnetohydrodynamics, and is defined by
\beqn
{\cal H}\equiv \int H^0 \sqrt{-g}\,d^3x\,, 
\label{Def_H}
\eeqn
where $H^0$ is the time component of the magnetic helicity four-current $H^\mu$, 
defined by 
\beqn
H^\mu\equiv -{1\over 2} \epsilon^{\mu\nu\alpha\beta} A_\nu F_{\alpha\beta}\,. 
\label{Def_Hvector}
\eeqn
Taking the covariant derivative of Equation (\ref{Def_Hvector}) yields 
\beqn
\nabla_\mu H^\mu=-{1\over 2} F^{*\mu\nu}F_{\mu\nu} \,. 
\eeqn
Thus, $\nabla_\mu H^\mu=0$ if $F_{\mu\nu}u^\nu=0$, i.e.,
$F^{*\mu\nu}=B^\mu u^\nu - B^\nu u^\mu$, and we confirm that the
magnetic helicity ${\cal H}$ is a conserved quantity in ideal
magnetohydrodynamics. For the present models, the total magnetic
helicity is explicitly written as
\beqn
{\cal H}={16\pi\over 3}(4\pi C_1)^2 L \int_0^R e^{\lambda-\nu} r^4 \psi_1^2\,dr \,, 
\eeqn
where the surface boundary condition (\ref{surface_bc}) has been
used. The dimensionless magnetic helicity, defined by ${\cal H}_M
\equiv {\cal H}/M^2$, is used when its numerical value is shown. The
magnetic helicity is a measure of the net twist of a magnetic-field
configuration. Thus, the magnetic helicity vanishes for purely
poloidal fields and for purely toroidal fields. Some experiments and
numerical computations show an interesting fact that the total
magnetic helicity is likely to be conserved even when the resistivity
cannot be ignored~\cite{ Braithwaite2004,Hsu2002}. If this fact is
retained for the neutron star formation process, the total magnetic
helicity has to be approximately conserved during its formation
process.

\section{Numerical results}

In this section, we present some numerical examples of stably
stratified stars composed of mixed poloidal-toroidal magnetic fields.
As one-parameter equations of state, we employ the following one,  
\beqn
P=\kappa\,\rho^{\displaystyle1+{1\over n}} \,, \label{eosa}\\
\varepsilon={1\over \Gamma-1}{P\over\rho}\,, \label{eosb}
\eeqn
where $\kappa$ and $n$ are the polytropic constant and index,
respectively, and $\Gamma$ denotes the adiabatic index, which is
defined by $\Gamma=(\partial\ln P/\partial \ln\rho)_S$.  $\kappa$,
$n$, and $\Gamma$ are constants and may be specified independently for
the construction of equilibrium stars, i.e. $\Gamma$ is not $1+1/n$ in
general. We define a general relativistic Schwarzschild discriminant
$A$ for non-magnetized spherical stars by
\beqn
A&\equiv&{1\over\rho h}{d \rho^*\over dr}-{1\over\Gamma}{1\over
P}{dP\over dr} \,,
\eeqn
where $\rho^*$ is the total energy density, defined by $\rho^*\equiv
\rho+\rho\varepsilon$.  Note that there is no unique definition
for the general relativistic Schwarzschild discriminant. However, in
the present context, only its sign matters.  For different definitions
of it, see, e.g., Refs.~\cite{throne66, mcdermott, ipser}.  Following
Ipser and Lindblom~\cite{ipser}, we employ a definition of the
Brunt--V{\"a}is{\"a}l{\"a} frequency $N \equiv \sqrt{-g A}$ with
\begin{equation}
g \equiv -e^{2(\nu-\lambda)} {1\over \rho h} {dP \over dr}. 
\end{equation}
If there is a region of $A>0$, the star has an unstably stratified
region and becomes convectively unstable there.  For the standard
stars whose density profile everywhere satisfies $d\rho/dr < 0$, the
condition of the stable stratification for the equations of state
(\ref{eosa}) and (\ref{eosb}) is given by
\beqn
\Gamma > {n+1\over n}\,. 
\eeqn
Note that for the isentropic case, defined by $\Gamma=(n+1)/n$, the
star is not stratified (marginally stable against convection), as
already mentioned.

Hereafter, we consider the $n=1$ case for simplicity, while we employ
$\Gamma=2$ and $2.1$: The $\Gamma=2$ models are non-stratified ones,
whose results are compared with those given in Ref.~\cite{Ioka2004} to
check our numerical results.  The choice of $\Gamma=2.1$ comes form
the following reason: As argued by Reisenegger and
Goldreich~\cite{reisenegger}, the Brunt--V{\"a}is{\"a}l{\"a} frequency
$N$ from the proton-neutron composition gradient inside neutron stars
is approximated by
\beqn
N \approx \sqrt{{x\over 2}} \sqrt{d\rho\over dP} \, g  \,,
\label{N_comp}
\eeqn
where $x$ denotes the ratio of the number densities of protons to
neutrons.  From Equation (\ref{N_comp}), it is found that the
$\Gamma=2.1$ models have Brunt--V{\"a}is{\"a}l{\"a} frequencies at the
stellar center similar to those of Reisenegger and Goldreich's models
with $x\approx 0.1$, which is a reasonable value for sufficiently cold
neutron stars. Note that the origins of the buoyancy in our models and
normal neutron stars are different.  In our models, the buoyancy
results from the entropy gradient not the composition gradient. 

First, we describe properties of the unperturbed stars.  In the
Newtonian framework, the profiles of a star like a density
distribution are independent of values of $\Gamma$ because Newtonian
gravity is determined only by the rest mass density and the pressure
is assumed to be a function of the rest mass density only. Thus, we
may calculate thermodynamical structures of the star like internal
energy for any value of $\Gamma$ after determining the structure of a
star.  In general relativity, by contrast, the profiles of the star do
depend on the value of $\Gamma$ because $P$ and $\varepsilon$ are a
source of gravity.  Thus, we have to recalculate stellar profiles in
general relativity whenever a value of $\Gamma$ is changed.

\begin{figure*}[b]
\epsfxsize=3.in
\hspace{-0.2cm}\epsffile{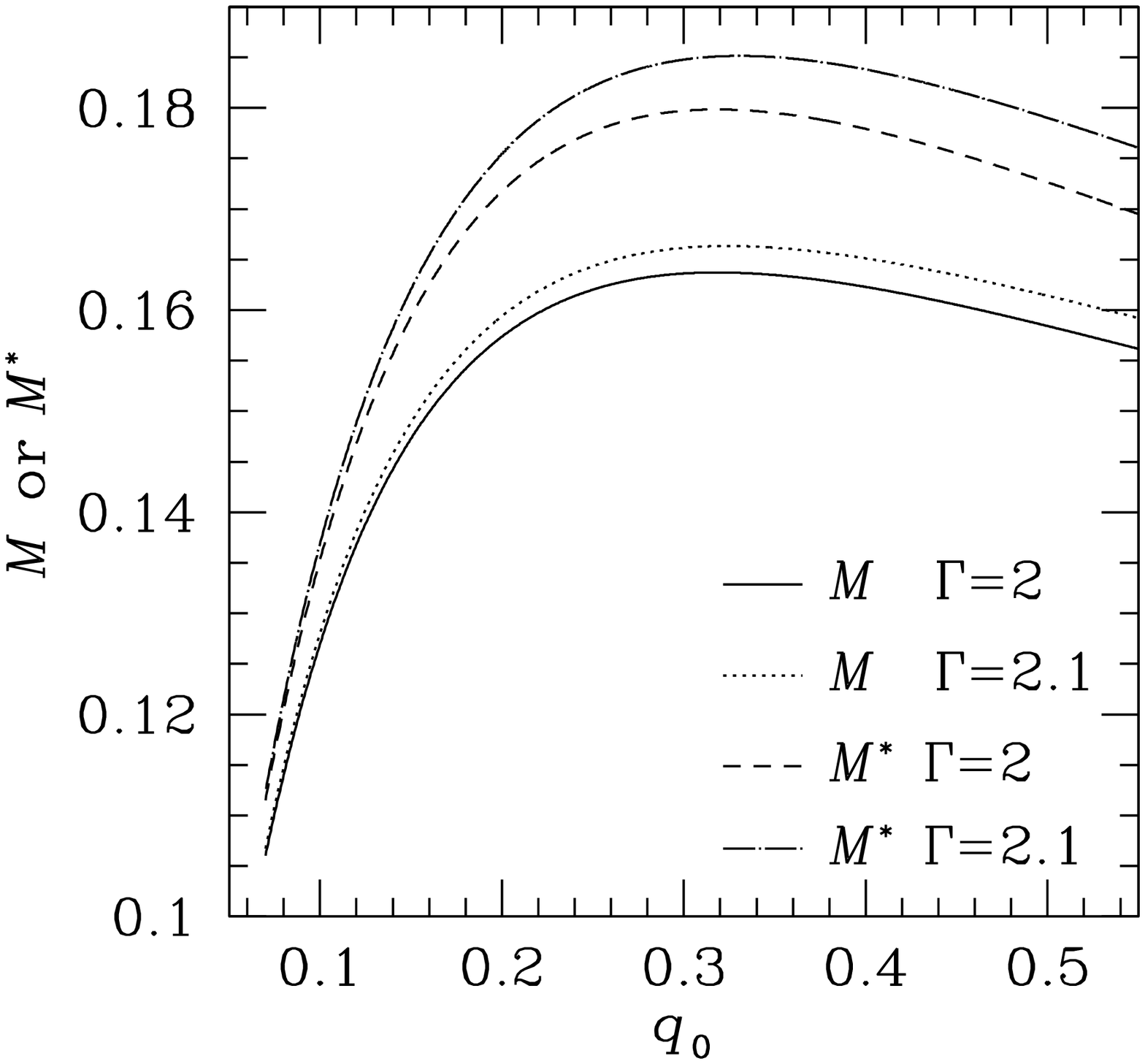}
\caption{Gravitational mass $M$ and baryon mass $M^*$, 
given as functions of the central density $q_0$. }
\label{f00}
\end{figure*}

\begin{figure*}[b]
\epsfxsize=3.in
\hspace{-0.2cm}\epsffile{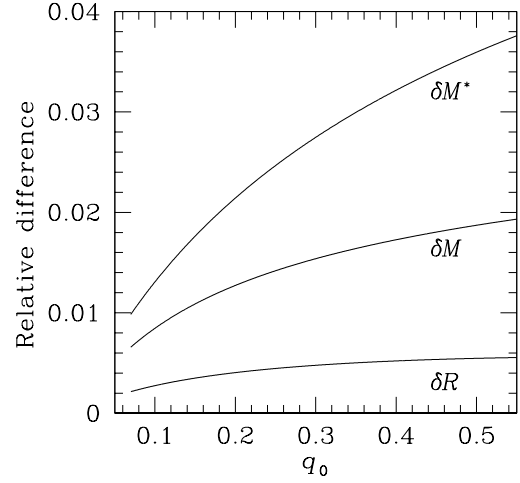}
\caption{Relative differences of  the gravitational mass $M$, 
the baryon mass $M^*$, and the radius $R$, given as a function of the
central density $q_0$.  Here, the relative difference of the physical
quantity $Q[\Gamma]$ is defined by $\delta Q \equiv
2(Q[2.1]-Q[2])/(Q[2.1]+Q[2])$. }
\label{f01}
\end{figure*}

In Figure \ref{f00}, the gravitational mass $M$ and the baryon mass
$M^*$ are plotted as functions of the central density of the star
$q_0\equiv\rho(r=0)$. Throughout this paper, we use units of 
$\kappa=1$ when showing numerical results.  This figure shows that
values of $M$ and $M^*$ for the $\Gamma=2.1$ models are larger than
those for the $\Gamma=2$ models for the same central density.  Figure
\ref{f01} plots relative differences between the $\Gamma=2.1$ and the
$\Gamma=2$ models in the gravitational mass $M$, the baryon mass
$M^*$, and the stellar radius $R$, as functions of the central density
$q_0$.  This shows that the radius of the star increases with
increasing $\Gamma$ if one keeps the central density constant.  The
reason is that an increase in $\Gamma$ reduces the total internal
energy of the star and leads to a decrease in an effective
gravitational attraction force.  Figure~\ref{f01} also shows that
$\delta M^* > \delta M > \delta R$ for all the values of $q_0$
calculated in the present study and that a typical value of the
relative difference in the baryon mass for a standard neutron star
model with $q_0\approx 0.2$ is within several percents.  Thus, we may
conclude that the $\Gamma=2.1$ model has properties quite similar to
those of the $\Gamma=2$ model, except for the stability against
convection: We emphasize again that the stratification condition of
the $\Gamma=2.1$ model is absolutely different from that of the
$\Gamma=2$ model.  Figures \ref{f00} and \ref{f01} show that the
gravitational mass, baryon mass, and radius of the stars tend to be
independent of $\Gamma$ as $q_0 \rightarrow 0$, i.e., in the Newtonian
limit. Henceforth, we will focus on the compact models with $M/R=0.1$
and $0.2$. The models having $M/R=0.2$ will be a reasonable model of a
neutron star.  In Table~\ref{table1}, some global and physical quantities
for the $\Gamma=2$ and $2.1$ models are tabulated.
\begin{table}[t]
\caption{Global and physical quantities for the background stars in
units of $\kappa=1$.}\label{table1}
\begin{tabular}{cccccc} \hline
~$\Gamma$~&~~$M/R$~~ & ~~~$q_0$~~~ & ~~~$M$~~~&~~~$M^*$~~~ &~$E_{\rm int}/|W|$~\\ 
\hline
$2.000$ & $0.1000$ & $0.07027$ & $0.1062$ & $0.1118$ & $0.4078$ \\
                & $0.2000$ & $0.25582$ & $0.1623$ & $0.1780$ & $0.5633$ \\
$2.100$ & $0.1000$ & $0.06983$ & $0.1066$ & $0.1124$ & $0.3706$ \\
                & $0.2000$ & $0.24893$ & $0.1643$ & $0.1820$ & $0.5104$ \\
\hline
\end{tabular}
\end{table}

Next we explore the magnetic-field profile around the background
spherical stars and its effects on the stellar structures. As already
mentioned in the previous section, the magnetic-field profile is
determined by solving an eigenvalue equation with respect to the
eigenvalue $L$, which is related to the toroidal magnetic-field
strength through $L=F_{12}\, \sin\theta e^{\nu-\lambda}
\Psi^{-1}$. Then, a discrete sequence of $L$ is allowed for the magnetized
stars satisfying the boundary conditions (\ref{psi_bs1}) and
(\ref{surface_bc}). Table~\ref{table2} lists a sequence of the
dimensionless eigenvalues $L$ in units of $R^{-1}$ for the first six
eigensolutions, where $L_i$ means the $i$-th eigenvalue $L$ satisfying 
$0 < L_1 < L_2 < L_3 < \cdots$.
\begin{table}[t]
\caption{Dimensionless low-order eigenvalues $L_i$ in units of $R^{-1}$.}\label{table2}
\begin{tabular}{ccccc} \hline
~$\Gamma$~&~Mode order~ & ~~$M/R=0.1000$~~ & ~~$M/R=0.2000$~~ \\ \hline
$2.000$&$R L_1$ & $5.792$ & $3.907$ \\ 
               &$R L_2$ & $8.315$ & $5.601$ \\
               &$R L_3$ & $10.80$ & $7.257$ \\
               &$R L_4$ & $13.26$ & $8.903$ \\ 
               &$R L_5$ & $15.71$ & $10.54$ \\
               &$R L_6$ & $18.16$ & $12.18$ \\ 
$2.100$&$R L_1$ & $5.787$ & $3.909$ \\ 
               &$R L_2$ & $8.316$ & $5.619$ \\
               &$R L_3$ & $10.80$ & $7.283$ \\
               &$R L_4$ & $13.26$ & $8.937$ \\ 
               &$R L_5$ & $15.71$ & $10.59$ \\
               &$R L_6$ & $18.16$ & $12.23$ \\
\hline
\end{tabular}
\end{table}
For the models with $\Gamma=2$, the values of $L$ in units of $R^{-1}$
are compared with those given in Table I of Ref.~\cite{Ioka2004} and
we confirm that our results are in excellent agreement with theirs.

Here, we should remark the following point. In the study of
Ref.~\cite{Ioka2004}, the meridional flow is in general
present. However, the meridional flow may be absent in the solutions
of Ref.~\cite{Ioka2004}.  This no meridional-flow limit is given by
the limit $|\tilde{C}| \rightarrow \infty$ with $\tilde{L}$ kept to be
constant, in the notation of Ref.~\cite{Ioka2004}. Even if we take
this limit, the magnetic field remains unchanged (see Equations (83)
-- (86) of Ref.~\cite{Ioka2004}) although the meridional flow vanishes
(see Equations (88) and (89) of Ref.~\cite{Ioka2004}). By using a
relation $h\,e^\nu = {\rm const.}$, which is satisfied for
non-magnetized ``isentropic'' spherical stars, we also show for
isentropic stars with no meridional flow that basic equations of
Ref.~\cite{Ioka2004}, (75), (94), (95), and (100) -- (103), become
equivalent to our basic equations, (\ref{GS2}), and (\ref{m0}) --
(\ref{h2}) (but note some differences in notation and definition of
physical quantities, e.g., their $\tilde{L}$ corresponds to our
$L$). Here, it should be emphasized that the parameter $\tilde{C}$
associated with the strength of the meridional flow does not appear in
the equations of Ref.~\cite{Ioka2004}, (75), (94), (95), and (100) --
(103), which fully determine the structure of magnetized stars with no
meridional flow.

Table~\ref{table2} shows that the dimensionless eigenvalues $R\times
L$ for the $\Gamma=2.1$ model are approximately equal to those for the
$\Gamma=2$ model. The same result is found in the flux function $\Psi$
if one regards $\Psi(r,\theta)$ as a function of $r/R$ not as $r$.
These facts imply that slight changes in $\Gamma$ do not affect values
of $R\times L$ and $\Psi(r/R,\theta)$ distributions.

\begin{figure*}[b]
\epsfxsize=1.8in
\leavevmode
\epsffile{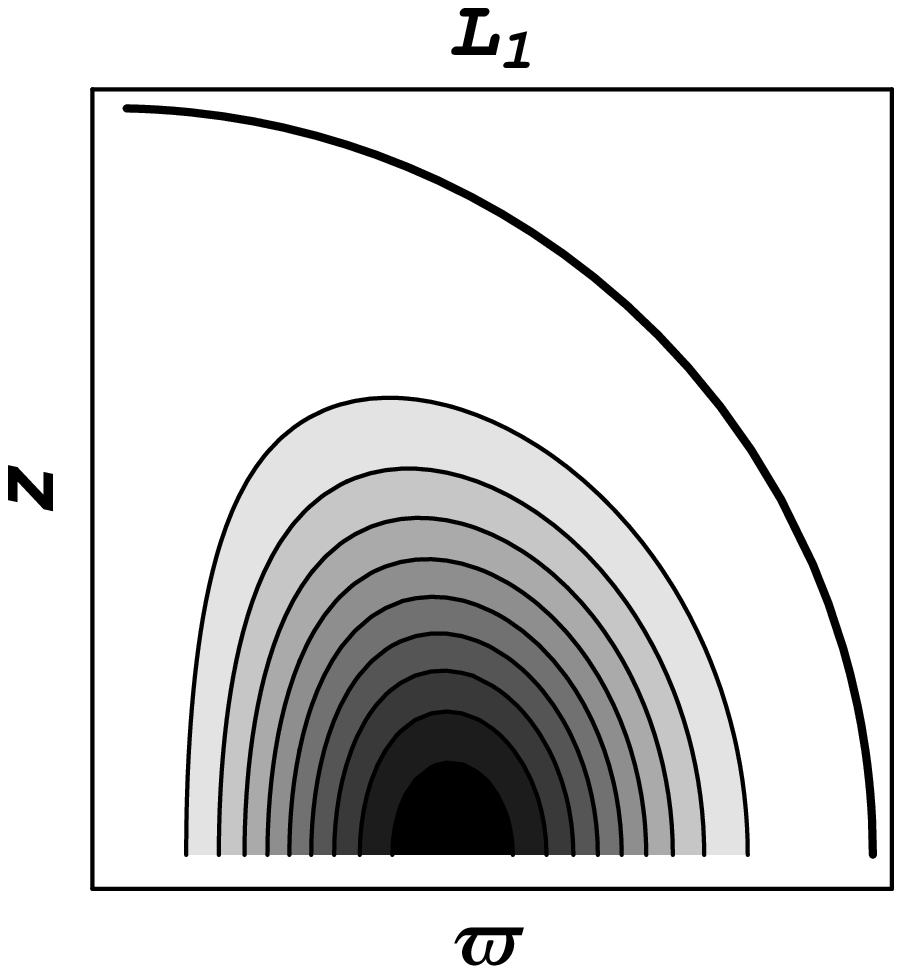}
\epsfxsize=1.8in
\leavevmode
\hspace{-0.2cm}\epsffile{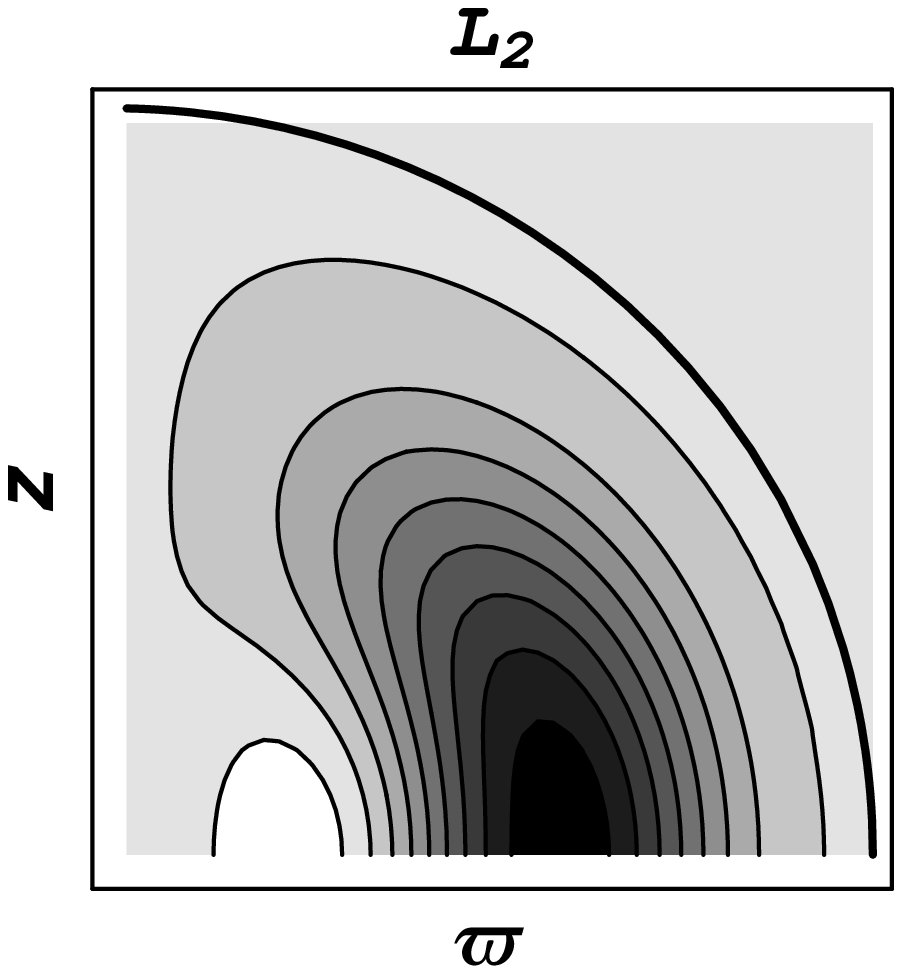}
\epsfxsize=1.8in
\leavevmode
\hspace{-0.2cm}\epsffile{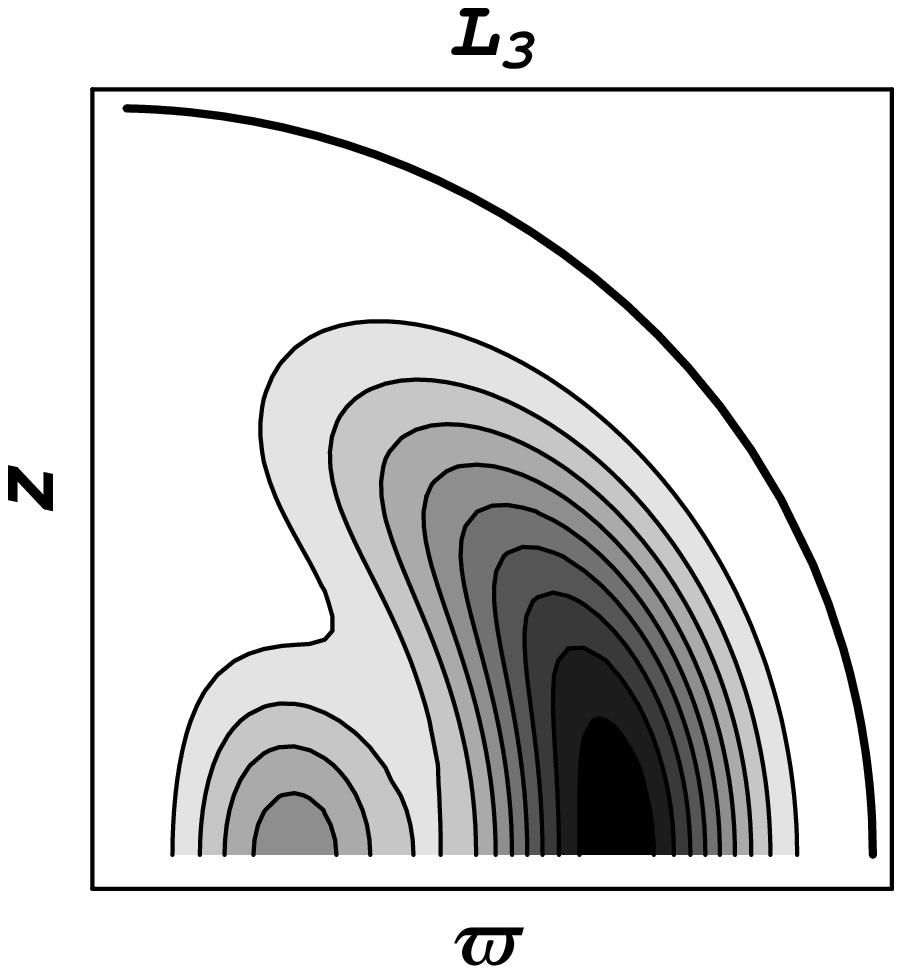}
\epsfxsize=1.8in
\leavevmode
\hspace{-0.2cm}\epsffile{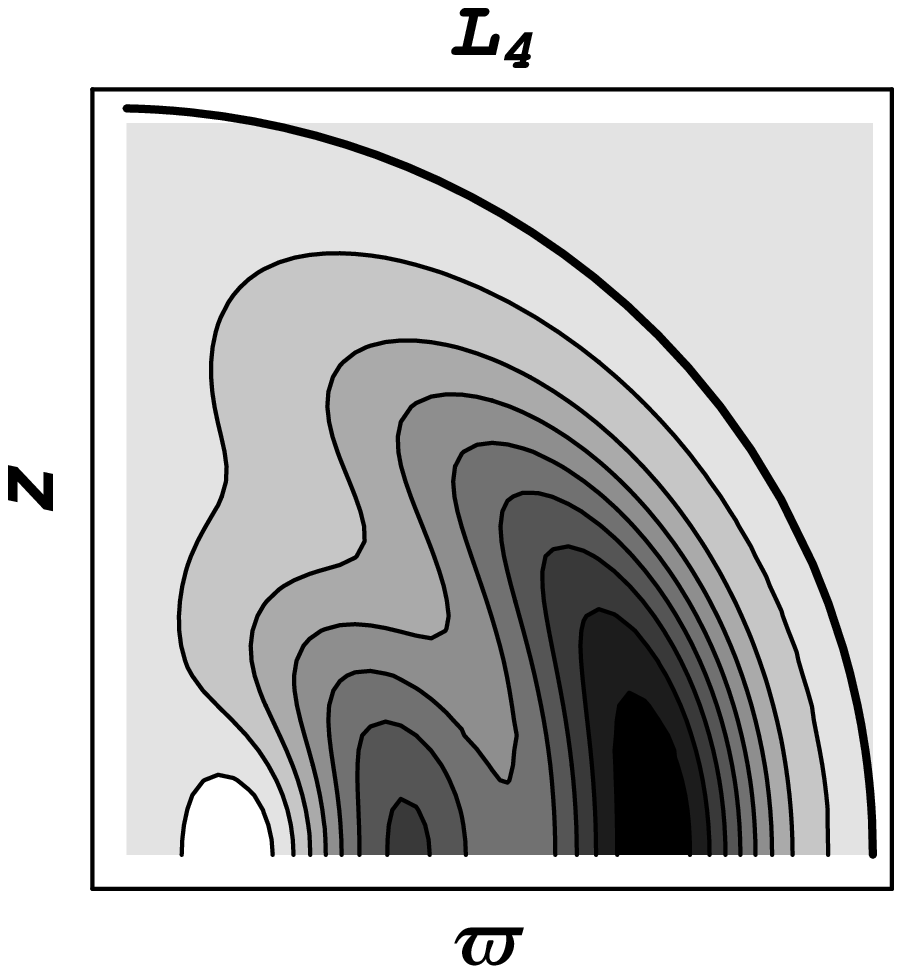}
\epsfxsize=1.8in
\leavevmode
\hspace{-0.2cm}\epsffile{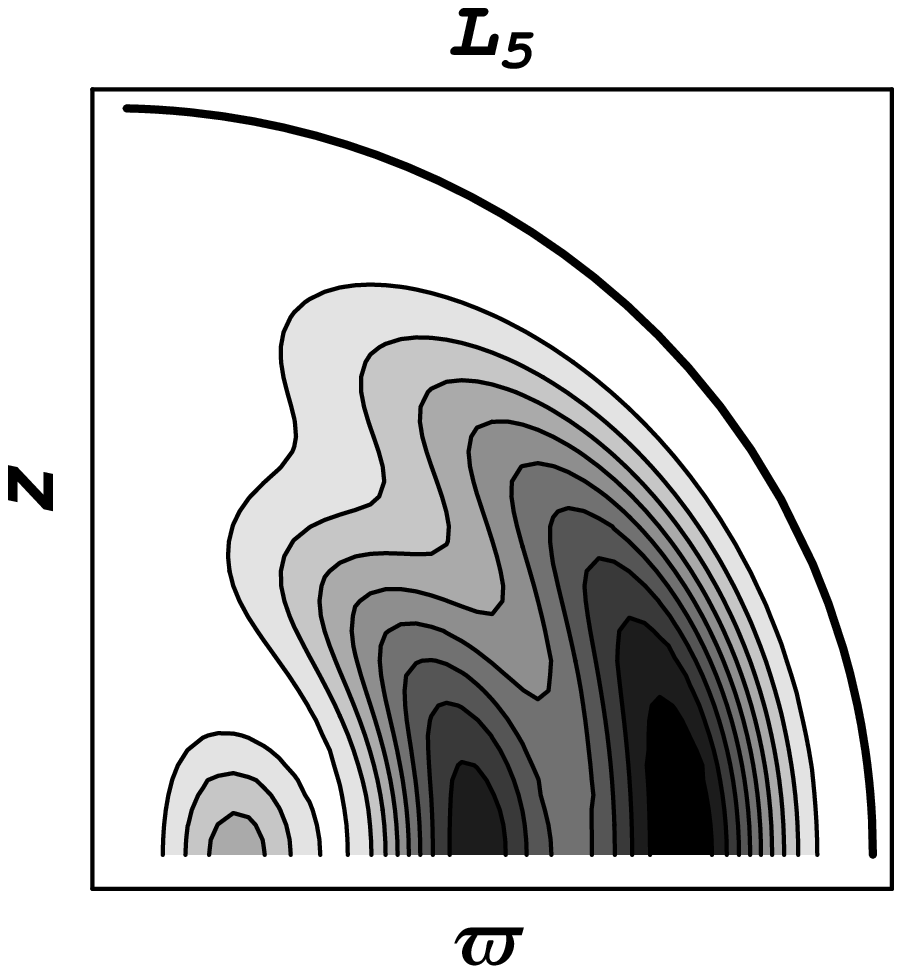}
\epsfxsize=1.8in
\leavevmode
\hspace{-0.2cm}\epsffile{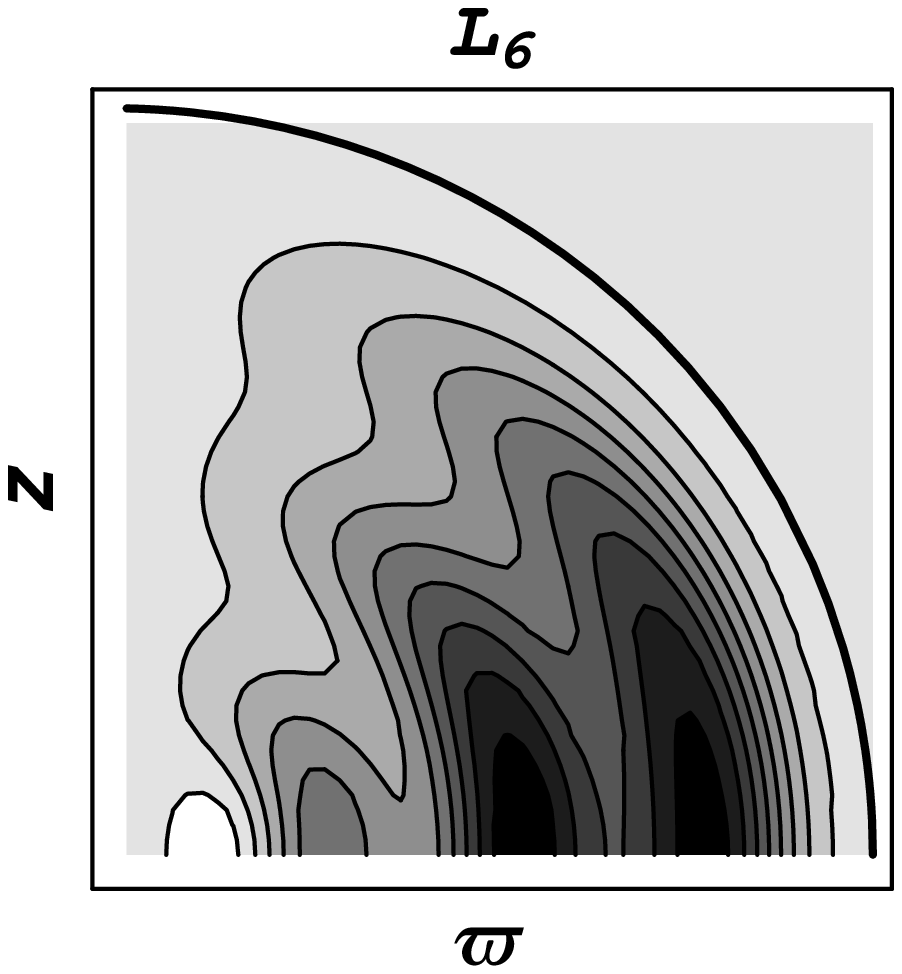}
\caption{Equi--$\Psi$ contours on the meridional cross section for the
  models characterized by $L=L_1$, $L_2$, $L_3$, $L_4$, $L_5$, and
  $L_6$. Here, $z$ and $\varpi$ are defined by $z\equiv r\cos\theta$
  and $\varpi=r\sin\theta$, respectively. The thick quarter circle
  shows the surface of the star, on which $\Psi=0$ is, by the boundary
  condition, required. The black and white regions, respectively,
  correspond to the maximum and minimum values of $\Psi$.  Since
  $\Psi$ vanishes outside the star, it can be seen that there is a
  negative region of $\Psi$ for the models with $L_2$, $L_4$, and
  $L_6$, while $\Psi$ is always positive for the models with $L=L_1$,
  $L_3$, and $L_5$.  }
\label{FIG1}
\end{figure*}

\begin{figure*}[b]
\epsfxsize=1.8in
\leavevmode
\epsffile{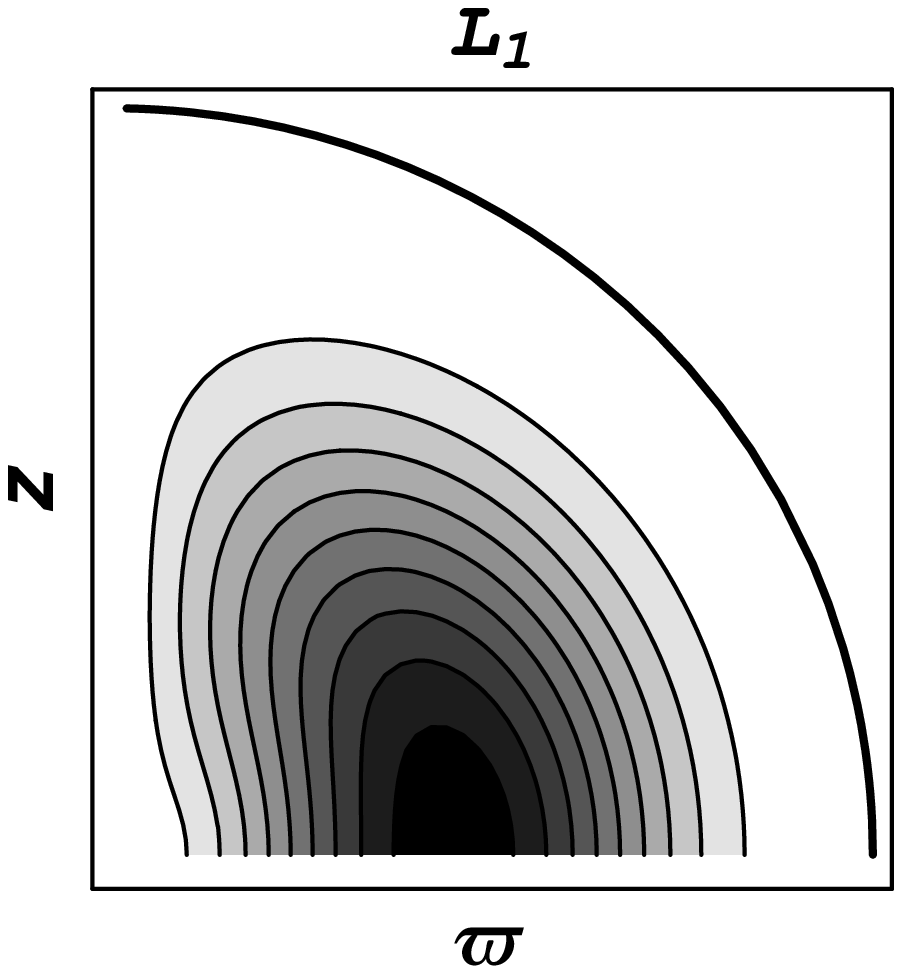}
\epsfxsize=1.8in
\leavevmode
\hspace{-0.2cm}\epsffile{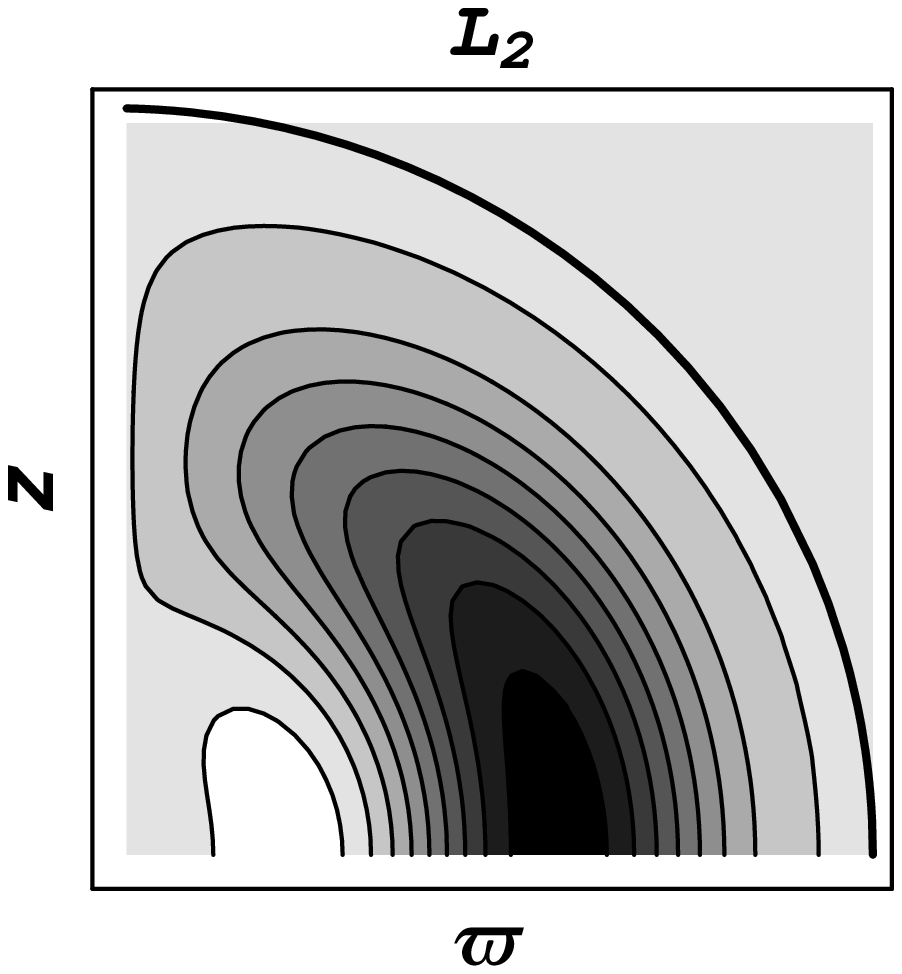}
\epsfxsize=1.8in
\leavevmode
\hspace{-0.2cm}\epsffile{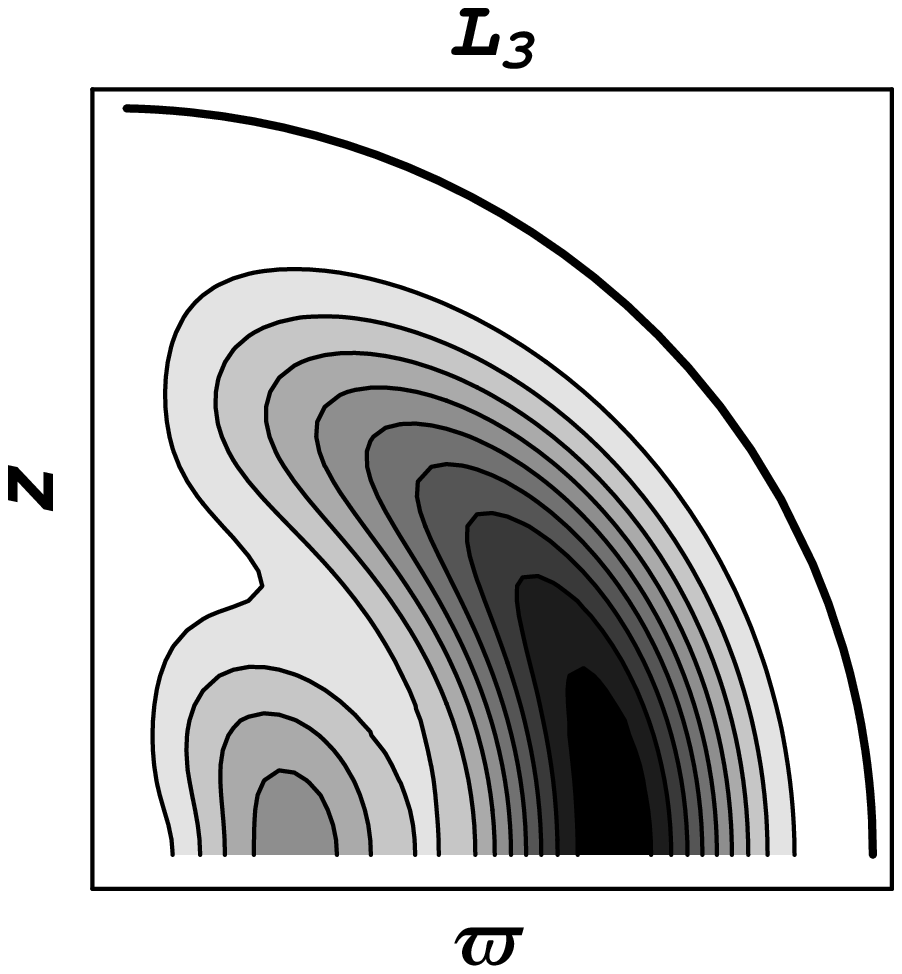}
\epsfxsize=1.8in
\leavevmode
\hspace{-0.2cm}\epsffile{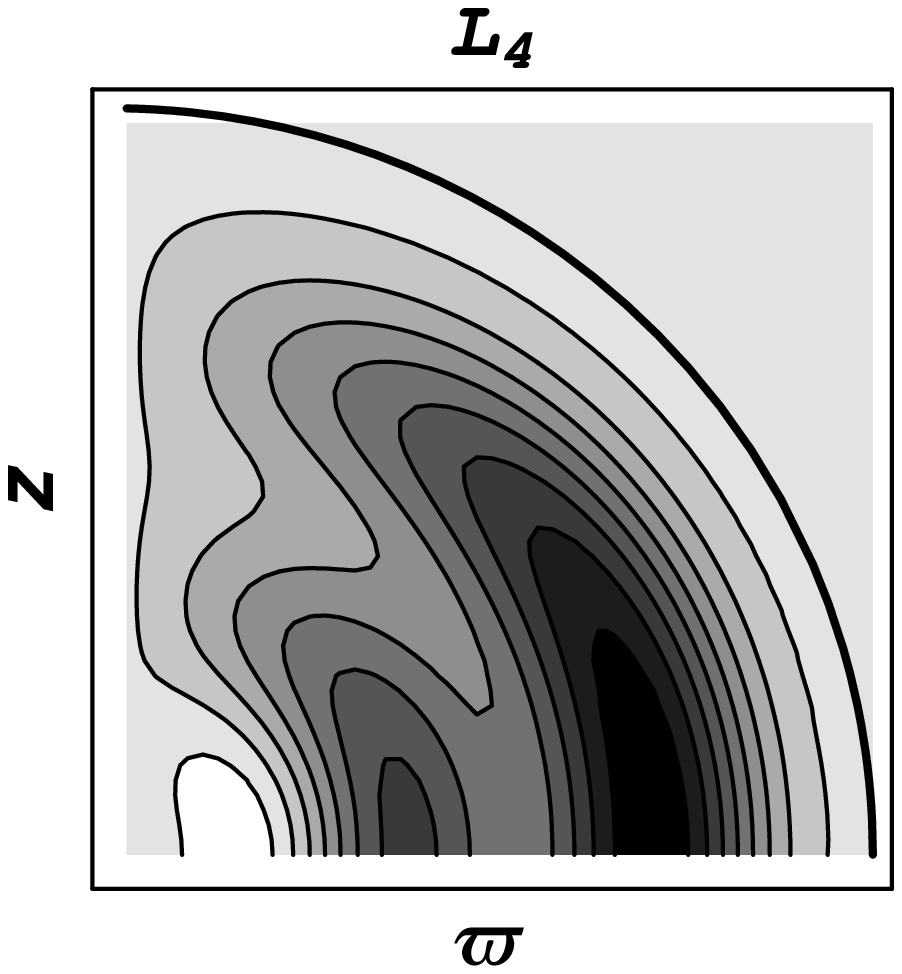}
\epsfxsize=1.8in
\leavevmode
\hspace{-0.2cm}\epsffile{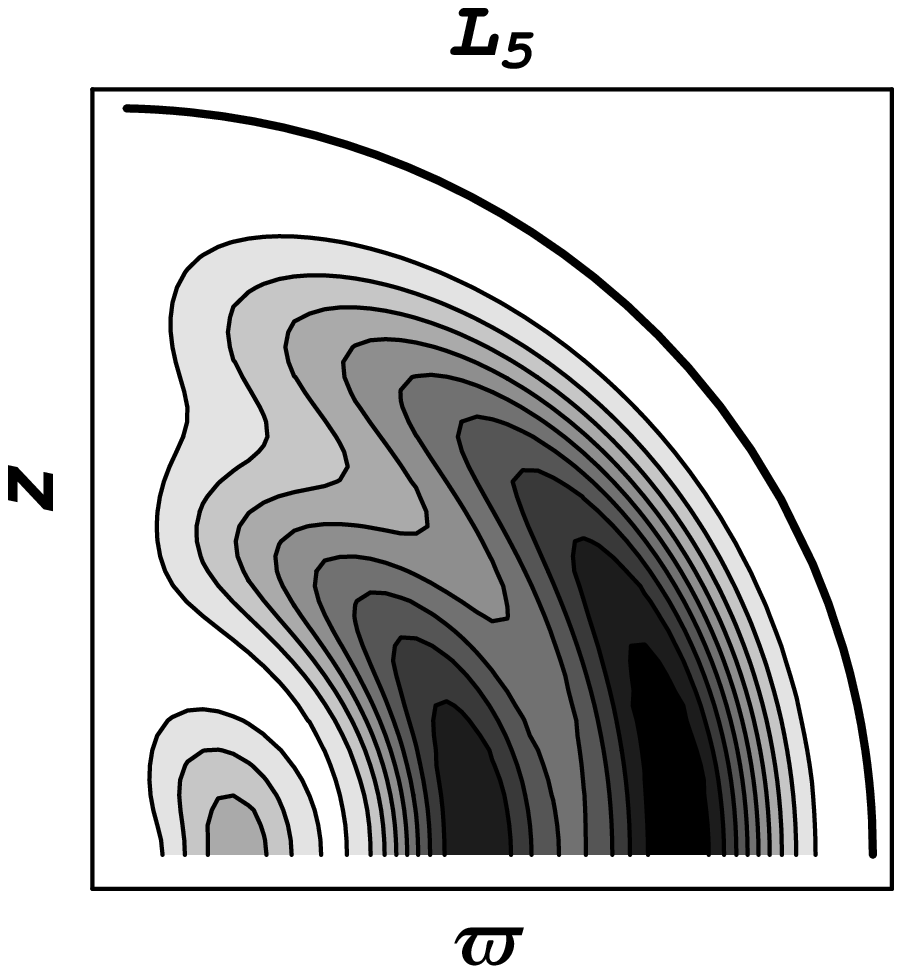}
\epsfxsize=1.8in
\leavevmode
\hspace{-0.2cm}\epsffile{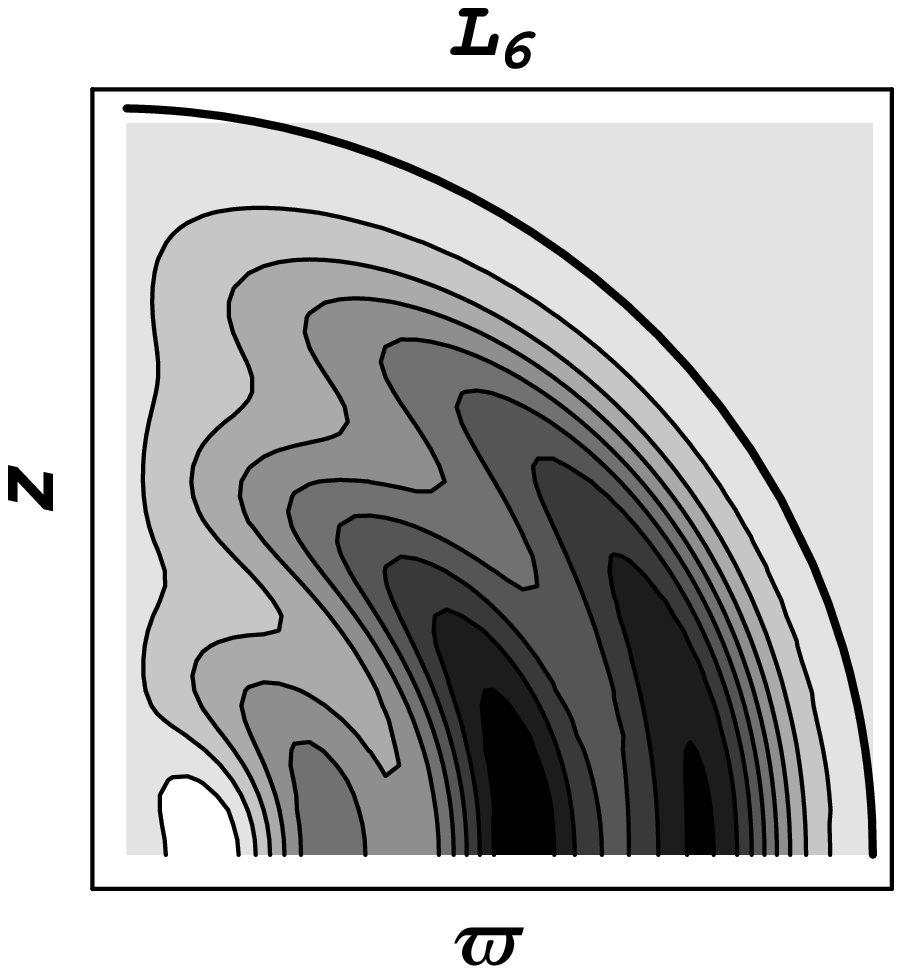}
\caption{ Equi--$F_{12}$ contours on the meridional cross section for
  the models characterized by $L=L_1$, $L_2$, $L_3$, $L_4$, $L_5$, and
  $L_6$. Here, $z$ and $\varpi$ are defined by $z\equiv r\cos\theta$
  and $\varpi=r\sin\theta$, respectively.  The thick quarter circle
  shows the surface of the star, on which $F_{12}=0$ is, by the
  boundary condition, required. The black and white regions,
  respectively, correspond to the maximum and minimum values of
  $F_{12}$.  Since $F_{12}$ vanishes outside the star, it can be seen
  that there is a negative region of $F_{12}$ for the models with
  $L_2$, $L_4$, and $L_6$, while $F_{12}$ is always positive for the
  models with $L=L_1$, $L_3$, and $L_5$.  }
\label{FIG2}
\end{figure*}

Figures~\ref{FIG1} and \ref{FIG2} display the profiles of magnetic
fields; eigensolutions of $\Psi$ and $F_{12}$ with $L=L_1$ -- $L_6$
for the spherical star with $M/R=0.2$ and $\Gamma=2.1$ (see the
corresponding eigenvalues in Table~\ref{table2}).  Figure \ref{FIG1}
shows how lines of the magnetic force on the meridional cross section
behave, because an equi-$\Psi$ line corresponds to a line of the
magnetic force.  These figures suggest that higher-order
eigensolutions have more non-uniform structures of the magnetic
fields.  Figure \ref{FIG1} also shows that there is a negative region
of $\Psi$ for the models with $L_2$, $L_4$, and $L_6$, while $\Psi$ is
everywhere non-negative for the models with $L=L_1$, $L_3$, and $L_5$.

For analyzing properties of magnetic-field profiles, it is helpful to
introduce an orthonormal tetrad component of the magnetic field,
$B_\mu$, given by
\beqn
B_{(t)}=0\,, \  B_{(r)}=-8\pi C_1\psi_1\cos\theta\,,  \nonumber \\
B_{(\theta)}=4\pi C_1e^{-\lambda} (r \psi_1'+2\psi_1)\sin\theta\,, \ \\
B_{(\phi)}=4\pi C_1e^{-\nu} L r \psi_1\sin\theta\,. \nonumber
\eeqn
Then, we can define that $B_c\equiv|B(r=0)|=8\pi C_1\psi_1(0)$. As
mentioned before, profiles of $B_{(\mu)}$ for the $\Gamma=2.1$ models
are very similar to those for the $\Gamma=2$ models. Their difference
cannot be visible if $B_{(\mu)}$ are plotted as functions of $r/R$ in
the same figure even though no figure is given in this paper. To show
the dimensionless total magnetic helicity of the magnetized stars,
$\cal{H}_M$, we follow Ioka and Sasaki
\cite{Ioka2004} and use a dimensionless parameter that represents
magnetic-field strength, defined by
\beqn
{\cal R}_M \equiv  {B_c^2R^4 \over 4 M^2} \,, 
\eeqn
which is as large as the ratio of the magnetic energy to the
gravitational energy.

\begin{table*}[t]
\begin{center}
\caption{Global and physical quantities characterizing the magnetized
stars with mixed poloidal-toroidal fields; the changes in the central
density $\Delta\rho_c$, the gravitational mass $\Delta M$, the
quadrupole moment $\Delta Q$, the mean radius $(\Delta r)_0$, the
ellipticity $e^*$, the magnetic helicity $\cal{H}_M$, the magnetic
energy $E_{\rm EM}$, and the ratio of the poloidal magnetic energy
$E_{\rm EM}^{(p)}$ to the total magnetic energy $E_{\rm EM}$. Here,
all the quantities are normalized to be non-dimensional, as given in
the first row.}\label{table3}
\begin{tabular}{cccccccccc} \hline
($\Gamma$, $M/R$) & $L$ &  ${\Delta\rho_c/q_0{\cal H}_M}$ & $\Delta M/M {\cal H}_M$ & 
$\Delta Q/M R^2 {\cal H}_M$ & $(\Delta r)_0/R {\cal H}_M$ &  $e^*/{\cal H}_M$ & ${\cal H}_M/{\cal R}_M$ 
& ${E_{\rm EM}/|W|\cal{H}_M}$ & $E_{\rm EM}^{(p)}/E_{\rm EM}$ \\ 
\hline
$(2, 0.2)$ & $L_1$ & $0.2545$ & $2.446\times 10^{-2}$ & $-6.163\times 10^{-3}$ & $-2.601\times 10^{-2}$ 
& $-1.505\times 10^{-2}$ & $2.556\times 10^{-1}$ & $0.1903$ & $0.3684$ \\
 & $L_2$ &$0.4063$ & $3.067\times 10^{-2}$ & $-1.875\times 10^{-2}$ & $9.007\times 10^{-2}$ 
& $-4.580\times 10^{-2}$ & $4.346\times 10^{-1}$ & $0.2222$ & $0.2785$ \\
 & $L_3$ &$0.5307$ & $3.634\times 10^{-2}$ & $-3.293\times 10^{-2}$ & $4.156\times 10^{-2}$ 
& $-8.043\times 10^{-2}$ & $1.153\times 10^{-1}$ & $0.2606$ & $0.2097$ \\
 & $L_4$ &$0.6736$ & $4.202\times 10^{-2}$ & $-4.719\times 10^{-2}$ & $7.387\times 10^{-2}$ 
& $-1.153\times 10^{-2}$ & $1.690\times 10^{-1}$ & $0.3000$ & $0.1607$ \\
 & $L_5$ &$0.7967$ & $4.780\times 10^{-2}$ & $-6.120\times 10^{-2}$ & $1.051\times 10^{-1}$ 
& $-1.495\times 10^{-1}$ & $7.632\times 10^{-2}$ & $0.3404$ & $0.1255$ \\
 & $L_6$ &$0.9303$ & $5.372\times 10^{-2}$ & $-7.490\times 10^{-2}$ & $1.353\times 10^{-1}$ 
& $-1.829\times 10^{-1}$ & $1.026\times 10^{-1}$ & $0.3818$ & $0.09996$ \\
$(2, 0.1)$ & $L_1$ & $-0.04212$ & $1.777\times 10^{-2}$ & $-1.957\times 10^{-2}$ & $9.053\times 10^{-2}$ 
& $-3.654\times 10^{-2}$ & $2.777\times 10^{-1}$ & $0.2488$ & $0.3541$  \\
 & $L_2$ &$0.1186$ & $2.223\times 10^{-2}$ & $-6.018\times 10^{-2}$ & $1.845\times 10^{-1}$ 
& $-1.123\times 10^{-1}$ & $5.802\times 10^{-1}$ & $0.3043$ & $0.2588$ \\
 & $L_3$ &$0.1856$ & $2.650\times 10^{-2}$ & $-1.036\times 10^{-1}$ & $2.797\times 10^{-1}$ 
& $-1.935\times 10^{-1}$ & $1.349\times 10^{-1}$ & $0.3612$ & $0.1913$ \\
 & $L_4$ &$0.3061$ & $3.074\times 10^{-2}$ & $-1.467\times 10^{-1}$ & $3.733\times 10^{-1}$ 
& $-2.738\times 10^{-1}$ & $2.208\times 10^{-1}$ & $0.4188$ & $0.1444$ \\
 & $L_5$ &$0.3825$ & $3.507\times 10^{-2}$ & $-1.886\times 10^{-1}$ & $4.649\times 10^{-1}$ 
& $-3.521\times 10^{-1}$ & $9.250\times 10^{-2}$ & $0.4777$ & $0.1116$ \\
 & $L_6$ &$0.4846$ & $3.953\times 10^{-2}$ & $-2.295\times 10^{-1}$ & $5.547\times 10^{-1}$ 
& $-4.284\times 10^{-1}$ & $1.338\times 10^{-1}$ & $0.5378$ & $0.08816$ \\
$(2.1,0.2)$ & $L_1$ & $0.2227$ & $2.417\times 10^{-2}$ & $-6.186\times 10^{-3}$ & $-1.747\times 10^{-2}$ 
& $-1.511\times 10^{-2}$ & $2.571\times 10^{-1}$ & $0.1912$ & $0.3694$ \\
 & $L_2$ &$0.3599$ & $3.047\times 10^{-2}$ & $-1.892\times 10^{-2}$ & $2.124\times 10^{-2}$ 
& $-4.622\times 10^{-2}$ & $4.146\times 10^{-1}$ & $0.2245$ & $0.2810$ \\
 & $L_3$ &$0.4711$ & $3.604\times 10^{-2}$ & $-3.340\times 10^{-2}$ & $5.790\times 10^{-2}$ 
& $-8.158\times 10^{-2}$ & $1.146\times 10^{-1}$ & $0.2632$ & $0.2120$ \\
 & $L_4$ &$0.6008$ & $4.162\times 10^{-2}$ & $-4.800\times 10^{-2}$ & $9.407\times 10^{-2}$ 
& $-1.172\times 10^{-1}$ & $1.638\times 10^{-1}$ & $0.3029$ & $0.1627$ \\
 & $L_5$ &$0.7115$ & $4.730\times 10^{-2}$ & $-6.233\times 10^{-2}$ & $1.292\times 10^{-1}$ 
& $-1.523\times 10^{-1}$ & $7.549\times 10^{-2}$ & $0.3435$ & $0.1272$ \\
 & $L_6$ &$0.8324$ & $5.311\times 10^{-2}$ & $-7.635\times 10^{-2}$ & $1.631\times 10^{-1}$ 
& $-1.865\times 10^{-1}$ & $9.981\times 10^{-2}$ & $0.3851$ & $0.1014$ \\
$(2.1, 0.1)$ & $L_1$ & $-0.03900$ & $1.852\times 10^{-2}$ & $-1.952\times 10^{-2}$ & $8.932\times 10^{-2}$ 
& $-3.643\times 10^{-2}$ & $2.777\times 10^{-1}$ & $0.2490$ & $0.3545$ \\
 & $L_2$ &$0.1208$ & $2.319\times 10^{-2}$ & $-6.015\times 10^{-2}$ & $1.832\times 10^{-1}$ 
& $-1.122\times 10^{-1}$ & $5.680\times 10^{-1}$ & $0.3050$ & $0.2597$ \\
 & $L_3$ &$0.1884$ & $2.763\times 10^{-2}$ & $-1.038\times 10^{-1}$ & $2.783\times 10^{-1}$ 
& $-1.938\times 10^{-1}$ & $1.343\times 10^{-1}$ & $0.3620$ & $0.1921$ \\
 & $L_4$ &$0.3091$ & $3.205\times 10^{-2}$ & $-1.470\times 10^{-1}$ & $3.719\times 10^{-1}$ 
& $-2.745\times 10^{-1}$ & $2.177\times 10^{-1}$ & $0.4196$ & $0.1451$ \\
 & $L_5$ &$0.3861$ & $3.655\times 10^{-2}$ & $-1.892\times 10^{-1}$ & $4.634\times 10^{-1}$ 
& $-3.531\times 10^{-1}$ & $9.192\times 10^{-2}$ & $0.4786$ & $0.1122$ \\
 & $L_6$ &$0.4885$ & $4.120\times 10^{-2}$ & $-2.302\times 10^{-1}$ & $5.531\times 10^{-1}$ 
& $-4.298\times 10^{-1}$ & $1.321\times 10^{-1}$ & $0.5388$ & $0.08865$ \\
\hline
\end{tabular}
\end{center}
\end{table*}

Table~\ref{table3} lists global physical quantities characterizing the
magnetized stars with mixed poloidal-toroidal fields; the changes in
the central density $\Delta\rho_c$, the gravitational mass $\Delta M$,
the quadrupole moment $\Delta Q$, the mean radius $(\Delta r)_0$, the
ellipticity $e^*$, the magnetic helicity $\cal{H}_M$, the magnetic
energy $E_{\rm EM}$, and the ratio of the poloidal magnetic energy
$E_{\rm EM}^{(p)}$ to the total magnetic energy $E_{\rm EM}$. In this
table, the results for the first six eigensolutions are shown and all
the quantities are normalized to be non-dimensional, as given in the
first row.  By comparing the results shown in Table~\ref{table3} with
those in Table 2 of Ref.~\cite{Ioka2004}, we can again check the
reliability of our results; ours are in agreement with theirs for the
$\Gamma=2$ models in an acceptable level. 

Although there are slight numerical differences between them, there is
no qualitative difference between the results of the $\Gamma=2.1$ and
$\Gamma=2$ models (see Table \ref{table3}). Thus, basic properties of
the magnetized star described in Ref.~\cite{Ioka2004} hold for the
present $\Gamma=2.1$ models, although the convective stability is
different and also any meridional flow is absent in the present
models.  For the magnetized-star sequences, characterized by the fixed
baryon mass and magnetic helicity, shown in Table~\ref{table3}, the
total gravitational mass and the total magnetic energy increase with
the mode number $i$ (or the eigenvalue $L$). This property is
reasonable due to the following reason: As found in Figures \ref{FIG1}
and \ref{FIG2}, higher-order eigensolutions, characterized by a lager
eigenvalue, have more non-uniform structures of the magnetic fields,
which can usually store larger magnetic-field energy. From
Table~\ref{table3}, we also find that all the models obtained in this
study are toroidal-magnetic-field dominant and that the value of
$E_{\rm EM}^{(p)}/E_{\rm EM}$ decreases as the mode number $i$
increases.  The later property is well explained by the fact that $L$
is interpreted as the ratio of the toroidal magnetic-field strength to
the poloidal magnetic-field one, and it increases as the mode number
$i$ increases. The magnetic hoop stress around the symmetry axis due
to the toroidal magnetic field tends to make the star prolate like a
rubber belt fastening a waist of the star. This is consistent with the
facts that all the stars obtained in this study have negative
ellipticity $e^*$, i.e., the star is prolate, and that the degree of
the quadrupole deformation measured by $|e^*|$ becomes more pronounced
for higher-order solutions, as shown in Table~\ref{table3}.

\section{Discussion: The stability of the stars}

Checking the stability is an important issue to be explored after
obtaining an equilibrium state of a magnetized star, because unstable
solutions lose their physical meaning in the sense that they are not
realized in nature. For the present models, as observed in Sec. V, the
gravitational mass and the total electromagnetic energy increase with
the mode number $i$ of the eigensolution, if we consider the
equilibrium sequences for a given baryon mass and a magnetic helicity.
This result is quite important and interesting due to the following
reason. If the total baryon mass and magnetic helicity are conserved
during the formation process of neutron stars, as discussed before, it
is likely that the final state of the magnetic fields characterized by
the arbitrary functions of the flux function (\ref{Def_f_Om}) --
(\ref{Def_f_Gam}) and the surface boundary condition
(\ref{surface_bc}) is the lowest-order eigensolution characterized by
the smallest eigenvalue $L=L_1$ because it has the lowest
gravitational mass and total electromagnetic energy among the
equilibrium solutions characterized by the fixed baryon mass and
magnetic helicity. We therefore conjecture that all the high-order
solutions are unstable because there is an equilibrium state with
energy lower than theirs and that the solutions associated with the
lowest eigenvalue $L=L_1$ can be stable if there is a stable solution
among the present models.  Another important fact is that for the
magnetic-field profile characterized by $L=L_1$, their magnetic energy
is most equally divided into the poloidal and the toroidal magnetic
energies among the eigensolutions. This property is consistent with
the conjecture for stable magnetic configurations given by the linear
analyses; stable magnetized stars contain both poloidal and toroidal
components with comparable magnetic energies.

\begin{figure*}[b]
\epsfxsize=3.in
\hspace{-0.2cm}\epsffile{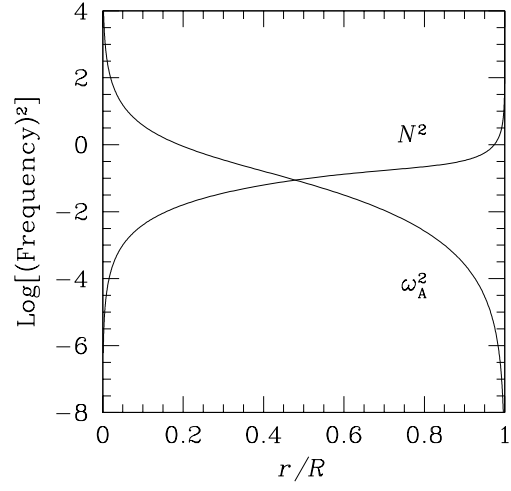}
\caption{Squares of the Brunt--V{\"a}is{\"a}l{\"a} frequency $N^2$ and
  the Alfv{\'e}n frequency $\omega_A^2$, given as functions of the
  dimensionless radius $r/R$.  Here, the Alfv{\'e}n frequency is
  defined by $\omega_A \equiv\sqrt{ B^\mu B_\mu/((4\pi\rho h + B^\mu
    B_\mu)r^2)}$ and evaluated on the equatorial plane
  ($\theta=\pi/2$), and the strength of the magnetic fields are
  determined by the condition $E_{\rm EM}/|W|=2.5\times 10^{-2}$. }
\label{f04}
\end{figure*}

Most of the magnetized-star models constructed in the framework of
general relativity so far were non-stratified, and therefore,
marginally stable against convection. Those non-stratified models are
in general highly unstable against the magnetic buoyancy in the
vicinity of the stellar surface, in the presence of magnetic
fields. For strongly magnetized stars like magnetars, as shown by
Kiuchi et al.~\cite{kiuchib}, the magnetic buoyancy instability
induces a convective motion near the surface of the star and fully
destroys initially coherent magnetic fields inside the star.  To
stabilize this magnetic-buoyancy instability, the stratification with
the strength sufficient to overcome the magnetic buoyancy are
necessary as a stabilizing agent.  This stabilization effect prevails
in non-rotating diffusion-less stars with purely toroidal magnetic
fields as argued by Acheson \cite{acheson1978}.  When $N^2 \gg
\omega_A^2>0$, his dispersion relation (see Equation (3.20) of
Ref.~\cite{acheson1978}) has four solutions, $\omega \approx \pm
k_\theta N(k^2_r+k^2_\theta)^{-{1\over 2}}$ and $\omega \approx \pm m
\omega_A$, where $\omega$ and $\omega_A$ mean the oscillation and the
Alfv{\'e}n frequencies, respectively, and $k_r$, $k_\theta$, and $m$
denote the vertical, horizontal, and azimuthal wave numbers,
respectively. These four solutions are composed of propagating waves;
the former is an internal gravity wave and the latter an Alfv{\'e}n
wave. We therefore confirm that there is no magnetic-buoyancy
instability as long as $N^2 \gg \omega_A^2>0$.  In Figure~\ref{f04},
we plot squares of the Brunt--V{\"a}is{\"a}l{\"a} frequency $N^2$ and
the Alfv{\'e}n frequency $\omega_A^2$ for a $\Gamma=2.1$ model
characterized by $M/R=0.2$ and $L=L_1$ as functions of the
dimensionless radius $r/R$.  Here, the Alfv{\'e}n frequency is
evaluated on the equatorial plane and defined by $\omega_A
\equiv\sqrt{ B^\mu B_\mu/((4\pi\rho h + B^\mu B_\mu)r^2)}$, and the
strength of the magnetic fields is determined by the condition $E_{\rm
  EM}/|W|=2.5\times 10^{-2}$, which corresponds to very strong
magnetic fields $\approx 10^{16}$~G for a typical neutron star.  This
figure shows that the Brunt--V{\"a}is{\"a}l{\"a} frequency, $N$, is
much larger than the Alfv{\'e}n frequency $\omega_A$ in the region of
$r > 0.5 R$ for the models with $L=L_1$. Thus, we can predict that
this model is stable against the magnetic buoyancy  
vicinity of the stellar surface.
It should be noted that
the physical origin of the anti-buoyancy force in our models is
different from that in real neutron stars, as mentioned before,
because the former comes from the entropy gradient and the later
mainly from the composition gradient (see, e.g.,
Ref.~\cite{reisenegger}).  Magnetic fields, however, do not care the
origin of the anti-buoyancy force, but do whether stars are stably
stratified or not.  

As pointed out by Acheson \cite{acheson1979}, the magnetic hoop stress
caused by the strong toroidal magnetic fields governs dynamics of the
perturbation near the center of the star. This fact may be confirmed
from Figure~\ref{f04}, which shows $N \ll\omega_A$ near the center of
the star. Thus, the stratification is not helpful near the center of
the star in the presence of the magnetic instability.  For the central
part of the star, as mentioned before, the presence of the poloidal
magnetic fields having comparable strength with the toroidal ones will
suppress the pinch-type instabilities. By solving linear perturbation
equations around magnetized star models in the framework of Newtonian
dynamics, Lander and Jones showed that this suppression of the
magnetic instability indeed occurs, although the presence of the
poloidal magnetic fields leads another instability associated with
themselves \cite{lander12} (see, also, Refs.~\cite{bonanno,bonanno2}).
Thus, the results of their numerical simulation lessen the possibility
that the pinch-type magnetic instabilities are completely removed for
the stars containing the mixed poloidal-toroidal magnetic fields with
comparable strength. Although Lander and Jones's results obviously
conflict with those by Braithwaite and his collaborators
\cite{Braithwaite2004,Braithwaite2006,Braithwaite2009,duez10}, we have
so far had no definite conclusion to this controversy.  One important
difference between their analyses is in the treatment of the
resistivity of the matter; Lander and Jones employed the ideal
magnetohydrodynamics approximation, whereas Braithwaite and his
collaborators took into account the resistivity.  This might be a key
to the solution of this problem. Toward a definite answer to this
stability problem, we need further studies on the stability of the
magnetized stars.

As such a study, we plan to perform a stability analysis of the
present magnetized star models by general relativistic
magnetohydrodynamics simulations like those done by Kiuchi et
al.~\cite{kiuchia,kiuchib}. The present stably stratified models
characterized by $L=L_1$ satisfy the following conditions; they have
both poloidal and toroidal magnetic fields with comparable strength to
suppress the hoop-stress instability inside the star, and in addition,
the fluid constructing the magnetized star is stably stratified with
strength sufficient to overcome the magnetic buoyancy near the stellar
surface. Thus, it is obvious that some major magnetic instabilities
will be reduced in the present magnetized star models. This will make
the stability problem more tractable.

As discussed in Kiuchi et al.~\cite{kiuchi,kiuchia}, it is reasonable
to expect that the toroidal component of the magnetic fields is much
larger than the poloidal component inside the neutron star at least
soon after its birth. The reason for this is that the winding of
poloidal magnetic fields caused by a rapid and differential rotation
during the core collapse would create large toroidal fields. Most of
general relativistic magnetized star models obtained in numerical
computations so far can, however, have toroidal magnetic fields much
weaker than poloidal ones \cite{colaiuda,ciolfia,ciolfib}. Their
minimum ratio of the poloidal magnetic field energy to the total
magnetic energy is $\approx 0.92$, which is much larger than those of
the present models.  (For magnetized star models in the framework of
the Newtonian dynamics, see, e.g., Ref.~\cite{haskell}). Their
magnetic fields are composed of the mixed poloidal--toroidal twisted
torus magnetic fields inside the star and nearly dipolar magnetic
fields outside the star. They may look quite plausible for the
magnetosphere of neutron stars. However, weak toroidal magnetic-field
strength seems to be unlikely for the strongly magnetized neutron
stars.  Although the magnetic field vanishes outside the star for the
present models, which is quite unrealistic, the present models would
give a reasonable inside structure of the strongly magnetized neutron
stars because of their large toroidal magnetic-field strength. For
obtaining more realistic models composed of a stably stratified fluid,
basic equations given in Equations (\ref{Def_Psi}) -- (\ref{current})
may be employed as far as ideal magnetohydrodynamics and the
barotropic equations of state are employed.

\section{Summary}

We constructed the magnetized stars composed of a stratified
fluid in the framework of general relativity. By assuming ideal
magnetohydrodynamics and employing a barotropic equation of state, we
first derive basic equations for describing stably stratified
stationary axisymmetric stars containing both poloidal and toroidal
magnetic fields.  As sample models, the magnetized star considered by
Ioka and Sasaki \cite{Ioka2004} are modified to the ones stably
stratified. The resulting models have both poloidal and toroidal
magnetic fields with comparable strength. The magnetized stars newly
constructed in this study are believed to be 
more stable than the existing relativistic models because they have
both poloidal and toroidal magnetic fields with comparable strength,
and  the magnetic buoyancy instability near the surface of the
star, which can be stabilized by the stratification, are
suppressed.

\acknowledgments
S.Y. thanks U. Lee for fruitful discussions.  This work was supported
by Grant-in-Aid for Scientific Research (22740178, 21340051, 24244028, 24540245)
and Grant-in-Aid on Innovative Area (20105004), and HPCI Stragetic
Program in Japan MEXT.

\end{document}